\documentstyle[12pt,aps,epsfig]{revtex}
\draft
\makeatother

\begin{document}

\title{Inter-scale correlations as measures of CMB Gaussianity}

\author{Serge Winitzki}

\address{Department of Physics and Astronomy, Case Western Reserve
University, Cleveland, OH 44106-7079.}

\author{Jiun-Huei Proty Wu}

\address{Department of Astronomy, University of California, Berkeley, CA
94720-3411.}

\maketitle
\begin{abstract}
A description of CMB temperature fluctuations beyond the power spectrum is
important for verifying models of structure formation, especially in view
of forthcoming high-resolution observations. We argue that higher-order
statistics of inter-scale correlations, because of their low cosmic
variance, may be effective in detecting non-Gaussian features in the CMB.
Inter-scale correlations are generically produced in defect-based models of
structure formation. We analytically study properties of general
higher-order cumulants of Fourier components of homogeneous random fields
and design a new set of statistics suitable for small-scale data analysis.
Using simulated non-Gaussian fields, we investigate the performance of the
proposed statistics in presence of a Gaussian background and pixel noise,
using the bispectrum as the underlying cumulant. Our numerical results
suggest that detection of non-Gaussian features by our method is reliable
if the power spectrum of the non-Gaussian components dominates that of the
Gaussian background and noise within at least a certain range of accessible
scales.

\end{abstract}

\section{Introduction}

The fluctuations in the cosmic microwave background (CMB) radiation first
reliably measured by the COBE \cite{COBE} are one of today's most
cosmologically important measurements (see \cite{Barreiro} for a recent
review). The future satellite missions MAP \cite{MAP} and PLANCK
\cite{PLANCK} are expected to provide high-resolution, low-noise CMB data
that will strongly constrain current theories of structure formation. One
of the key differences between the competing theories is an approximately
Gaussian distribution of density fluctuations produced in most inflationary
theories versus the generically non-Gaussian inhomogeneities in scenarios
based on topological defects. A precise statistical analysis of the CMB
fluctuations is needed to distinguish non-Gaussian signatures of cosmic
defects \cite{CMB-strings} and small deviations from Gaussianity in
inflationary universe due to generic effects such as a non-linear evolution
\cite{Bisp2,Gangui99}, gravitational lensing \cite{CMB-lensing}, and
higher-order couplings during recombination \cite{CMB-2nd}. The power
spectrum, which is the statistic easiest both to predict theoretically and
to extract from data, is by definition insensitive to non-Gaussian
correlations. For this reason, and especially in view of forthcoming
observations, detection of non-Gaussian features in the CMB is an important
task.

Since all theories give only statistical predictions of the CMB, one's
conclusions from a single observation of the microwave sky are by necessity
probabilistic. Many criteria of Gaussianity have been proposed in the
literature and applied to the available data, mostly yielding results
consistent with the Gaussian hypothesis \cite{Kogut}. Although some claims
of finding a non-Gaussian signal in COBE have been advanced recently
\cite{joao-ng1,NFS,PandoWav,joao-ng2}, not all of them are equally
persuasive and some have been ascribed to data contamination and noise in
Refs.~\cite{Tegmark,PandoNot}. Apart from these problems, the usefulness of
any CMB statistic is fundamentally limited by cosmic variance. Therefore we
would like to look for statistical descriptors of non-Gaussian signal that
have a naturally low variance.

Cosmic variance manifests itself differently in real space and in the
Fourier space where the CMB temperature fluctuations are usually decomposed
into modes $a_{lm}$ using the spherical harmonic expansion
\begin{equation}
\frac{\Delta T}{T}\left( \theta ,\phi \right) =\sum
_{l,m}a_{lm}Y_{lm}\left( \theta ,\phi \right) .
\end{equation}
The power spectrum estimator
\begin{equation}
C_{l}=\frac{1}{2l+1}\sum _{m=-l}^{l}\left| a_{lm}\right| ^{2}
\end{equation}
is increasingly more precise at smaller angular scales (higher $l$). In
other words, the cosmic variance in the Fourier space is shifted away from
small scales toward large scales. One would therefore expect any statistics
of Fourier modes at very small scales to be relatively free of cosmic
variance. On the other hand, real space data is correlated and estimators
based on correlations of $\Delta T\left( \theta ,\phi \right) $ are equally
affected by cosmic variance throughout all angular scales. (The recently
introduced wavelet-based statistics \cite{Hob1,PandoWav} can be regarded as
occupying the middle ground between the real and the Fourier space
descriptors. See also \cite{Heavens:peak} for an example of a low-variance
real-space statistic which is however strongly power-spectrum dependent.)

This motivates us to consider small-scale Fourier modes in hopes of
obtaining a sensitive statistic. Previous research \cite{FanBardeen}
suggests that individual Fourier modes are likely to be nearly Gaussian
distributed because of the central limit theorem. Therefore we intend to
investigate inter-scale correlations in Fourier space for the role of
non-Gaussian indicators. A non-Gaussian signal may manifest itself as a
correlation between Fourier modes $a_{lm}$ of different scales $l$; this
extra correlation must be of third order or higher because homogeneity
requires $\left\langle a_{lm}a^{*}_{l'm'}\right\rangle \propto \delta
_{ll'}\delta _{mm'}$. The third-order correlator of the Fourier modes,
called the ``bispectrum'', has been extensively studied in the literature
\cite{Bisp1,Bisp4,Heavens:bisp,Bisp3,Bisp2} and recently applied to the
COBE data \cite{joao-ng1,joao-ng2}. It was shown that the bispectrum
carries signatures from cosmic defects, as well as from non-linearities of
evolution in inflationary models. Defects generically produce non-Gaussian
features also in higher-order correlations, as we shall show.

We would like to reformulate the problem of finding the cross-correlation
between some chosen scales $l$ and $l^{\prime }$ as a Gaussianity test on
a suitable distribution. Consider for simplicity a random field in flat
(two-dimensional) space, decomposed into Fourier modes $a_{{\mathbf k}}$.
We may regard the set of mode values $a_{{\mathbf k}},\, a_{{\mathbf k}'}$
at the scales $\left| {\mathbf k}\right| $, $\left| {\mathbf k}'\right| $
as a sample of a two-variable distribution $\left\{ a,a'\right\} $ and
proceed to test that distribution for Gaussianity. A general way of testing
a distribution for Gaussianity is by using cumulants (see
e.g.~\cite{joao-c}). In our case we need to employ suitable multi-variable
cumulants, such as a third-order cumulant
\begin{equation}
\chi \left( k_{1},k'_{1},k_{2}\right) =\left\langle a_{{\mathbf
k}_{1}}a_{{\mathbf k}'_{1}}a_{{\mathbf k}_{2}}\right\rangle ,
\end{equation}
with $\left| {\mathbf k}_{1}\right| =\left| {\mathbf k}'_{1}\right| \neq
\left| {\mathbf k}_{2}\right| $. Such cumulants describe non-Gaussian
correlations between perturbations at different scales. Since the variance
of a cumulant estimator is decreased when the number of points in the
sample grows, we expect the statistic to be increasingly more sensitive on
smaller scales (larger $l$). We shall demonstrate for a Gaussian field that
Fourier space cumulant estimators of any order $n$, which we denote $\chi
\left( k_{1},...,k_{n}\right) $, at small scales (large $k_{i}$) are
statistically approximately independent and \emph{themselves} approximately
Gaussian distributed. We also find the variances of these cumulant
estimators of any order. This significantly simplifies the likelihood
analysis, since the variances are theoretically known and the estimators
can be normalized to have unit variances.

In other words, if the underlying CMB map were Gaussian, all quantities
$\chi \left( k_{1},...,k_{n}\right) $ for all scales $k_{i}$ form (after
normalization) a sample of independent realizations of a normal
distribution. One could test this hypothesis by choosing an appropriate
range of accessible scales $k_{i}$, combining all estimators $\chi \left(
k_{1},...,k_{n}\right) $ at these scales into one sample and computing the
first few cumulants $\omega _{j}$ (we take $j=1,...,5$) of that sample. If
the Gaussian hypothesis holds for the original distribution, the quantities
$\omega _{j}$ are in turn Gaussian distributed with known means and
variances. We now propose this as a Gaussianity test for the original map.
Likelihood analysis of the descriptors $\omega _{j}$ is at least as strong
as a simultaneous fit of all inter-scale cumulants and is potentially more
discriminating.

In this method, one is free to choose a subset of cumulants $\chi \left(
k_{1},...,k_{n}\right) $ and a subset of scales $k_{i}$ at which to
evaluate them. Since the statistical variance of the cumulant estimators is
decreased at smaller scales (larger $k_{i}$) but grows as $n!$ with the
order $n$ of the cumulants, the most promising results should come from the
lowest cumulants, such as the bispectrum ($n=3$), and at smallest available
scales. In this initial investigation we use only the bispectrum to build
the estimators $\omega _{n}$.

To evaluate the efficiency of the method, we use several simulated
non-Gaussian fields inspired by defect-based structure formation models.
One of the simulated models consists a superposition of fixed temperature
profiles centered on a random set of Poisson distributed points. The
advantage of this model is that for any shape of the profile and for any
distribution of the profile intensity and size, one can analytically obtain
the full generating functional of the resulting random field, which allows
(in principle) to evaluate any statistic. In particular, it is
straightforward to estimate the expected inter-scale correlations of
Fourier modes of this field. Another model we used is a simulated CMB map
from cosmic strings \cite{Proty}, superimposed on Gaussian background and
pixel noise. Our goal in all simulations was to find the levels of Gaussian
backgrounds at which the non-Gaussian signal is still detectable with our
statistic, given a certain level of noise.

The outline of the paper is as follows. In Sec.~\ref{sec:cumu} we define
the particular cumulants that describe inter-scale correlations of the
Fourier modes of a random field and show how we estimate them from a single
field realization. We derive the means and variances of these cumulant
estimators assuming Gaussian random field, as would be needed for
likelihood analysis, and show that the error bars decrease for large $l$,
as expected. Details of the calculations are given in Appendices
\ref{app:dc} and \ref{app:mc}. In Sec.~\ref{sec:ngm} we describe a model
non-Gaussian field made up of a random superposition of shapes,
analytically determine the expected non-vanishing cumulants, and
investigate the sensitivity of our criterion for that model in presence of
Gaussian backgrounds. The necessary formalism is developed in Appendix
\ref{app:gfs}. In Sec.~\ref{sec:num} we present numerical results obtained
with several simulated non-Gaussian maps using the lowest-order statistic
based on the bispectrum. The method is tested on maps of point sources,
random rectangles, and cosmic strings, mixed with Gaussian background and
noise. We give brief conclusions about the viability of the proposed
method.

\section{Cumulants in Fourier space}

\label{sec:cumu}For simplicity, in this section we consider random fields
on a plane; the results can be straightforwardly generalized to random
fields on a sphere, such as the CMB temperature perturbation $\Delta
T\left( \phi ,\theta \right) /T$. From the Fourier space viewpoint, a
random field $f\left( x\right) $ is a collection of random variables
(modes) $a_{{\mathbf k}}$,
\begin{equation}
a_{{\mathbf k}}=\frac{1}{\left( 2\pi \right) ^{2}}\int e^{-i{\mathbf
kx}}f\left( {\mathbf x}\right) d{\mathbf x}.
\end{equation}
Homogeneity of the random field (translation invariance in real space)
constrains the joint distribution of $\left\{ a_{{\mathbf k}}\right\} $,
namely the moments must satisfy
\begin{equation}
\label{eq:homgen}
\left\langle a_{{\mathbf k}_{1}}a_{{\mathbf k}_{2}}...a_{{\mathbf
k}_{n}}\right\rangle =0\textrm{ if }{\mathbf k}_{1}+...+{\mathbf k}_{n}\neq
0,
\end{equation}
and isotropy means that moments are invariant under rotations $R$ in the
Fourier space,
\begin{equation}
\label{eq:isogen}
\left\langle a_{{\mathbf k}_{1}}a_{{\mathbf k}_{2}}...a_{{\mathbf
k}_{n}}\right\rangle =\left\langle a_{R{\mathbf k}_{1}}a_{R{\mathbf
k}_{2}}...a_{R{\mathbf k}_{n}}\right\rangle .
\end{equation}
(In the spherical case, there is only one condition of invariance under
rotations of the sphere.) The task of testing Gaussianity of the random
field is now translated into checking whether all $a_{{\mathbf k}}$ are
jointly Gaussian distributed.

In general, one could describe the joint distribution of all Fourier modes
by a suitable generating  functional of moments,
\[
Z\left[ j\left( {\mathbf k}\right) \right] \equiv \sum _{n=0}^{\infty }\int
d{\mathbf k}_{1}...d{\mathbf k}_{n}\frac{j\left( {\mathbf k}_{1}\right)
...j\left( {\mathbf k}_{n}\right) }{i^{n}n!}\left\langle a_{{\mathbf
k}_{1}}a_{{\mathbf k}_{2}}...a_{{\mathbf k}_{n}}\right\rangle \]
\begin{equation}
\label{eq:fougenz}
=\left\langle \exp \left[ -i\int j\left( {\mathbf k}\right) a_{{\mathbf
k}}d{\mathbf k}\right] \right\rangle ,
\end{equation}
from which one recovers all the moments by functional differentiation,
\begin{equation}
\left\langle a_{{\mathbf k}_{1}}a_{{\mathbf k}_{2}}...a_{{\mathbf
k}_{n}}\right\rangle =i^{n}\frac{\delta ^{n}}{\delta j\left( {\mathbf
k}_{1}\right) ...\delta j\left( {\mathbf k}_{n}\right) }Z\left[ j\left(
{\mathbf k}\right) \right] _{j=0}.
\end{equation}
Here the functional argument $j\left( {\mathbf k}\right) $ is a
complex-valued function of ${\mathbf k}$ satisfying $j\left( -{\mathbf
k}\right) =j^{*}\left( {\mathbf k}\right) $. It is convenient to define the
general cumulants of the Fourier modes as quantities generated by the
logarithm of the functional of Eq.~(\ref{eq:fougenz}),
\begin{equation}
\label{eq:foucumz}
\tilde{C}^{\left( n\right) }\left( {\mathbf k}_{1},...,{\mathbf
k}_{n}\right) \equiv i^{n}\frac{\delta ^{n}}{\delta j\left( {\mathbf
k}_{1}\right) ...\delta j\left( {\mathbf k}_{n}\right) }\ln Z\left[ j\left(
{\mathbf k}\right) \right] _{j=0}.
\end{equation}
(The notation $\tilde{C}^{\left( n\right) }$ is chosen to be consistent
with Eq.~(\ref{c-f-gen}) of Appendix C while $C^{\left( n\right) }$ is
reserved for real-space cumulants.) For example, the distribution of modes
of a Gaussian random field with power spectrum $P\left( k\right) $ is
characterized by the generating functional
\begin{equation}
\label{eq:zgau}
Z_{G}\left[ j\left( {\mathbf k}\right) \right] =\exp \left[
-\frac{1}{2}\int P\left( k\right) \left| j\left( {\mathbf k}\right) \right|
^{2}d{\mathbf k}\right] ,
\end{equation}
and, as expected, all cumulants $\tilde{C}^{\left( n\right) }$ of order
$n\geq 3$ vanish for this distribution. It is easy to see that the general
cumulants $\tilde{C}^{\left( n\right) }$ satisfy homogeneity and isotropy
conditions similar to Eqs.~(\ref{eq:homgen})-(\ref{eq:isogen}). This is
because Eqs.~(\ref{eq:homgen})-(\ref{eq:isogen}) imply that the generating
functional $Z\left[ j\left( {\mathbf k}\right) \right] $ is invariant under
substitutions $j\left( {\mathbf k}\right) \rightarrow j\left( R{\mathbf
k}\right) $ and $j\left( {\mathbf k}\right) \rightarrow e^{i{\mathbf
kx}}j\left( {\mathbf k}\right) $ of its functional argument, and therefore
$\ln Z\left[ j\left( {\mathbf k}\right) \right] $, the generating
functional for cumulants, must also be rotation- and translation-invariant.
This, for instance, constrains the cumulants $\tilde{C}^{\left( n\right)
}\left( {\mathbf k}_{1},...,{\mathbf k}_{n}\right) $ to identically vanish
unless ${\mathbf k}_{1}+...+{\mathbf k}_{n}=0$.

Now we turn to statistics of $a_{{\mathbf k}}$ that are relevant for the
analysis of data coming from one observation. Since a measurement of one
realization of the random field gives only one set of values $\left\{
a_{{\mathbf k}}\right\} $, we are limited in the kinds of information about
the joint distribution of $\left\{ a_{{\mathbf k}}\right\} $ that we can
extracted without \emph{a priori} knowledge. For instance, we cannot
efficiently test Eqs.~(\ref{eq:homgen})-(\ref{eq:isogen}) for any
particular choice of $\left\{ {\mathbf k}_{i}\right\} $ because in each
case we would have only one value: $a_{{\mathbf k}_{1}}a_{{\mathbf
k}_{2}}...a_{{\mathbf k}_{n}}$. An inhomogeneous random field that has a
``wrong'' distribution of just one mode $a_{{\mathbf k}}$ cannot be
distinguished from a homogeneous and isotropic field on the basis of one
sample. (Of course, one would not expect such an artificial random field to
have physical relevance.)

Similarly, we cannot test the hypothesis that any two particular modes
$a_{{\mathbf k}_{1}}$ and $a_{{\mathbf k}_{2}}$ come from a jointly
Gaussian distribution if only one realization of these modes is available.
Clearly one can efficiently test a distribution for Gaussianity only if one
has a large enough number of samples of that distribution. Therefore, to
obtain any result at all concerning the Gaussianity of $a_{{\mathbf k}}$,
we must assume that several of the modes $a_{{\mathbf k}}$ come from the
same distribution. From the natural assumption of isotropy it follows that
the modes $a_{{\mathbf k}}$ within a ring of fixed scale $\left| {\mathbf
k}\right| =k$ are identically distributed, and thus the ring $\left|
{\mathbf k}\right| =k$ provides a sample that allows us to test their
distribution for Gaussianity. This is the assumption behind the bispectrum
test. Strictly speaking, a negative result of such a test would indicate
either a non-Gaussian or an anisotropic distribution. However, most
theories predict isotropy of CMB, and, assuming that all foregrounds are
adequately dealt with, we are much less interested in detecting anisotropy
or inhomogeneity than we are in finding non-Gaussian signals.

We are therefore motivated to assume homogeneity and isotropy and to regard
the modes $a_{{\mathbf k}}$ at a fixed scale $\left| {\mathbf k}\right| $
as independent samples of the same joint distribution of modes. This
assumption also allows us to investigate correlations between different
scales, which is the main interest of the present article. Consider two
rings of modes $a_{{\mathbf k}_{1,2}}$ at two fixed scales $k_{1}$ and
$k_{2}$. The values $a_{{\mathbf k}_{1,2}}$ along the two rings may be
regarded as independent realizations of the joint two-variable
distribution. Now we would like to test whether that distribution is
Gaussian in two variables.

A standard way to test a sample for Gaussianity is to compute cumulants of
various orders (see e.g. \cite{joao-c}): for a Gaussian distribution, the
cumulants of order $\geq 3$ vanish. Cumulants of a distribution of two
variables $\left( x,y\right) $ are quantities labeled by two indices,
e.g.~$\chi _{mn}$ has ``dimension'' $x^{m}y^{n}$; the general definition of
$\chi _{mn}$ can be given by
Eqs.~(\ref{eq:cumu-two-gen})-(\ref{eq:mom-two-gen}) of Appendix B. We are
interested in cumulants that describe cross-correlation between the two
variables, for instance a nontrivial third-order cross-cumulant is
\begin{equation}
\label{eq:chi12def}
\chi _{12}=\left\langle xy^{2}\right\rangle -\left\langle x\right\rangle
\left\langle y^{2}\right\rangle -2\left\langle xy\right\rangle \left\langle
y\right\rangle +2\left\langle x\right\rangle \left\langle y\right\rangle
^{2}.
\end{equation}
Similarly, one can define cumulants of a three-variable distribution
$\left( x,y,z\right) $, for example
\begin{equation}
\label{eq:chi111def}
\chi _{111}=\left\langle xyz\right\rangle -\left\langle xy\right\rangle
\left\langle z\right\rangle -\left\langle xz\right\rangle \left\langle
y\right\rangle -\left\langle yz\right\rangle \left\langle x\right\rangle
+2\left\langle x\right\rangle \left\langle y\right\rangle \left\langle
z\right\rangle .
\end{equation}

Given a sample of $N$ points $\left( x_{i},y_{i}\right) $ of a two-variable
distribution, one could estimate the cumulant $\chi _{12}$ directly by
evaluating the sample averages required by Eq.~(\ref{eq:chi12def}). As
shown in Appendix A, such estimators are generally not unbiased, and for a
Gaussian field they yield a non-zero expectation value of order $N^{-1}$,
as given by Eqs.~(\ref{chi1anse})-(\ref{chi1anso}) for one-variable and by
Eq.~(\ref{eq:chimnpq-ave}) for two-variable cumulants. This of course
presents no problem for likelihood analysis since this expectation value is
known for a Gaussian field and can be simply subtracted. We also derive the
expected variances of cumulant estimators first for one-variable cumulants
in Appendix A, and then for multivariable and Fourier space cumulants in
Appendix B. We show in Eqs.~(\ref{chi2ans}),
(\ref{eq:chimnqp-cov})-(\ref{eq:varcgen}) that the variance of a
multivariable cumulant estimator for a sample of $N$ simultaneous
realizations of a Gaussian distribution of $n$ variables $\left\{
x_{i}\right\} $ is always of order $N^{-1}$ and is given by
\begin{equation}
\textrm{var}\left[ \hat{\chi }_{l_{1}...l_{n}}\right]
=\frac{l_{1}!...l_{n}!}{N}\sigma _{1}^{2l_{1}}...\sigma
_{n}^{2l_{n}}+O\left( N^{-2}\right) ,
\end{equation}
where $\sigma _{i}$ are dispersions of the variables $x_{i}$. This
general result confirms and generalizes the formula obtained numerically in
Ref.~\cite{joao-c} for one-variable ($n=1$) cumulants. We see that since
the number $N$ of points in the sample is equal to the number of modes
$a_{{\mathbf k}}$ at the chosen scale $\left| {\mathbf k}\right| $,
the variance is indeed decreased for larger $\left| {\mathbf k}\right| $.

We could directly apply the cumulant technique to the Fourier modes by
taking e.g. $x\equiv a_{{\mathbf k}}$, $y\equiv a_{{\mathbf k}^{\prime }}$
in Eq.~(\ref{eq:chi12def}). The homogeneity constraint
Eq.~(\ref{eq:homgen}) suggests that of all possible cross-cumulants
$\tilde{C}^{\left( n\right) }\left( {\mathbf k}_{1},...,{\mathbf
k}_{n}\right) $, only those for which $\sum _{i}{\mathbf k}_{i}=0$ need to
be considered as possible non-Gaussian signals, all others being ``noise''
resulting from accidental inhomogeneity of the given sample. For instance,
the third-order cumulant relating two chosen scales $k_{1}$ and $k_{2}$
should be estimated from the set of triples $\left\{ a_{{\mathbf
k}_{1}},a_{{\mathbf k}^{\prime }_{1}},a_{{\mathbf k}_{2}}\right\} $ where
${\mathbf k}_{1}+{\mathbf k}_{1}^{\prime }+{\mathbf k}_{2}=0$ and $\left|
{\mathbf k}_{1}\right| =\left| {\mathbf k}'_{1}\right| =k_{1}$, $\left|
{\mathbf k}_{2}\right| =k_{2}$. (Clearly, the cross-cumulant of the scales
$k_{1}$ and $k_{2}$ can be nonzero only if $k_{2}\leq 2k_{1}$.) This
procedure is equivalent to regarding the triples $\left\{ a_{{\mathbf
k}_{1}},a_{{\mathbf k}^{\prime }_{1}},a_{{\mathbf k}_{2}}\right\} $ as
realizations of a \emph{three}-variable distribution for which we are
evaluating the cross-cumulant $\chi _{111}$, while in the notation of
Eq.~(\ref{eq:foucumz}), $\chi _{111}=\tilde{C}^{\left( 3\right) }\left(
{\mathbf k}_{1},{\mathbf k}_{1}^{\prime },{\mathbf k}_{2}\right) $. Below
we shall denote this inter-scale cumulant by $\chi _{111}\left(
k_{1},k_{2}\right) $. In Appendix B we derive general expressions for the
variances of such cross-cumulants. The sample size $N$ is determined by the
number of modes in the smallest of the rings $k_{1}$ and $k_{2}$, and the
variance of $\chi _{111}\left( k_{1},k_{2}\right) $ is
\begin{eqnarray}
\textrm{var}\left[ \chi _{111}\left( k_{1},k_{2}\right) \right] = &
N^{-1}\left[ P\left( k_{1}\right) \right] ^{2}P\left( k_{2}\right)  & \quad
(2k_{1}<k_{2}),\nonumber \\ \textrm{var}\left[ \chi _{111}\left(
k_{1},k_{2}\right) \right] = & 2N^{-1}\left[ P\left( k_{1}\right) \right]
^{2}P\left( k_{2}\right)  & \quad (2k_{1}=k_{2}).\label{eq:varchi21}
\end{eqnarray}

We show in Appendix B that Fourier mode cumulant estimators for a Gaussian
field are always uncorrelated Gaussian random variables, up to terms of
order $N^{-2}$. The Gaussianity hypothesis will hold for a distribution if
its cumulants estimated from the sample of $N$ data points fall within
their sample variances. The likelihood analysis in our case consists of
computing the chosen cumulant estimators (such as $\chi _{111}$) for
various pairs $\left( k_{1},k_{2}\right) $ of scales and normalizing them
to their theoretical Gaussian variances; the power spectrum needs to be
estimated beforehand from the same map. The resulting set of normalized
cumulant estimators contains as many numbers as there are pairs of
different scales $\left( k_{1},k_{2}\right) $ in the map and should be a
set of normally distributed, independent random values, if the underlying
map is Gaussian. We propose to test this by computing the first few
cumulants of that set, which is at least equivalent to joint estimation of
Gaussianity of the inter-scale cumulants for all pairs of scales.

\section{A non-Gaussian model}

\label{sec:ngm}To estimate the sensitivity of the Fourier
cross-correlations to non-Gaussian signal, we use the model of randomly
superimposed shapes similar to that of Ref.~\cite{Scherrer}. The random
field $f\left( x\right) $ is constructed as a superposition of ``seeds'',
i.e.~fixed profiles $s\left( x\right) $ located at random points $x_{i}$ in
the $2$-dimensional space. In addition, the profiles are attenuated, scaled
and rotated randomly,
\begin{equation}
f\left( x;\{x_{k},\nu _{k},\lambda _{k},R_{k}\}\right) =\sum _{k}\nu
_{k}s\left( \lambda _{k}^{-1}R_{k}\left( x-x_{k}\right) \right) .
\end{equation}
Here the number of seeds $n$ and seed positions $x_{i}$ are randomly
chosen in some predefined way; $\nu _{k}$ are attenuation factors, $\lambda
_{k}$ are scale factors, and $R_{k}$ are rotations, all drawn randomly for
each seed out of their corresponding distributions $p_{\nu }\left( \nu
\right) d\nu $ and $p_{\lambda }\left( \lambda \right) d\lambda $
(rotations are uniformly distributed in the rotation group). Such random
fields are physically relevant since the shapes could come from point
sources or individual topological defects, randomly positioned in the sky.

We would like to find out whether our criterion can distinguish a certain
level of this non-Gaussian signal in presence of Gaussian background or
noise. If we construct the full generating functional of the non-Gaussian
random field analytically, we can superimpose a Gaussian background on it
and obtain analytic predictions for the sensitivity of our criterion.

Nice properties of the Poisson distribution make it possible to obtain the
full generating functional for the random field with Poisson distributed
seeds. It is given by Eq.~(\ref{lnz-gen-r}) of Appendix C,
\begin{equation}
\ln Z\left[ J\left( x\right) \right] =-n_{c}+n_{c}\int dx_{0}dR\,
p_{\lambda }\left( \lambda \right) d\lambda \, p_{\nu }\left( \nu \right)
d\nu \, e^{-i\nu \int s\left( \lambda ^{-1}R\left( x-x_{0}\right) \right)
J\left( x\right) dx}\: .
\end{equation}
Here $n_{c}$ is the mean density of seeds. The non-Gaussian components
of the distribution are read directly from the generating functional which
shows that, in general, non-Gaussian cumulants of all orders are present,
cf.~Eq.~(\ref{c-f-gen}). As expected from the central limit theorem, the
relative magnitude of non-Gaussian components decays when the number
density of seeds grows.

A more general result relating the generating functionals of the seed
distribution and of the resulting random field is expressed by
Eqs.~(\ref{eq:zfzaux}) and (\ref{eq:zfzsans}). It shows that the generating
functional of Eq.~(\ref{lnz-gen-r}) could be expressed analytically because
the Poisson distribution of seeds is described by an analytic generating
functional. In more complicated cases or in situations where only a few
first seed correlations $\xi \left( x_{1},...,x_{n}\right) $ are known, the
full generating functional will not be available but we can still use these
general equations to obtain, accordingly, the first few cumulants of the
resulting random field $f\left( x\right) $.

These results are consistent with the formalism of Ref.~\cite{Scherrer}
where instead of scaling, rotation, or attenuation of seed profiles, a
distribution of seed masses $m_{i}$ was considered, the shape profile
$s\left( x-x_{i};m_{i}\right) $ being a function of the seed mass. Our
formalism may be generalized to treat the shape profile $s\left(
x;X_{i}\right) $ as an arbitrary function of $X_{i}$ where the
``coordinate'' parameter $X_{i}$ includes the position of the seed as well
as its mass, scaling etc., and the seed distribution must be specified with
respect to this generalized parameter. An analogue of
Eq.~(\ref{eq:zfzsans}) will hold also in this case; the generating
functional of the random field will be simply related to the generating
functional of seeds.

Similar results hold also for the generating functional $Z_{NG}\left[
j\left( {\mathbf k}\right) \right] $ for the Fourier components of the
non-Gaussian random field $f\left( x\right) $ {[}cf.~Eq.~(\ref{eq:zjk}){]}.
We can add to $f\left( x\right) $ a Gaussian background with a known power
spectrum $P_{G}\left( k\right) $; the background can be described by a
generating functional of Eq.~(\ref{eq:zgau}). The new generating functional
will be the product of the two and the cumulants will be the sums of the
two sets of cumulants. Since the Gaussian field has vanishing higher-order
cumulants, the only change will be in the power spectrum, which becomes
\begin{equation}
P\left( k\right) =P_{NG}\left( k\right) +P_{G}\left( k\right) .
\end{equation}
Consider one cumulant estimator $\tilde{C}^{\left( n\right) }\left(
k_{1},...,k_{n}\right) $ for a certain selected set of scales $k_{i}$.
Since its variance is inversely proportional to the appropriate powers of
$P\left( k\right) $ as given by Eq.~(\ref{eq:varcgen}), while its
expectation value is unchanged after adding the Gaussian background, the
sensitivity of the cumulant estimator will be diminished by the factor
\begin{equation}
\left( 1+\frac{P_{G}\left( k_{1}\right) }{P_{NG}\left( k_{1}\right)
}\right) ^{\frac{1}{2}}...\left( 1+\frac{P_{G}\left( k_{1}\right)
}{P_{NG}\left( k_{1}\right) }\right) ^{\frac{1}{2}}.
\end{equation}
This suggests that the sensitivity of cumulant estimators is unaffected on
scales where the Gaussian background is negligible compared to the
non-Gaussian signal, and will be proportionately decreased otherwise.

\section{Numerical results}

\label{sec:num}In this section we present the numerical procedures and
results of testing the sensitivity of our method.

\subsection{The scheme}

\label{cmb-ng2-the-scheme}

The procedure for testing a CMB map for Gaussianity using the Fourier space
cumulant criterion described in the previous sections can be summarized as
follows:

\begin{enumerate}
\item Fourier transform the map to obtain the modes $\left\{ a_{{\mathbf
k}}\right\} $ (this step would be unnecessary for data obtained from
interferometers). Estimate the power spectrum $P\left( k\right) $.
\item Divide the Fourier domain into rings of fixed moduli $k=\left|
{\mathbf k}\right| $. For each radius $k_{1}$ going from $k_{\min
}=0.1k_{\max }$ to $k_{\max }$, where $k_{\max }$ is the highest accessible
wavenumber, and for each radius $k_{2}$ going from $k_{\min }$ to $2k_{1}$,
we take a pair of modes $a_{{\mathbf k}_{1}}$ and $a_{{\mathbf
k}_{1}^{\prime }}$ from the ring of radius $k_{1}$, that is with $\left|
{\mathbf k}_{1}\right| =\left| {\mathbf k}_{1}^{\prime }\right| =k_{1}$,
and a mode $a_{{\mathbf k}_{2}}$ from the second ring of radius $k_{2}$.
The angle $\theta $ between the vectors ${\mathbf k}_{1}$ and ${\mathbf
k}_{1}^{\prime }$ is determined from
\begin{equation}
k_{2}=2k_{1}\cos \frac{\theta }{2}.
\end{equation}
The values of the two modes on the first ring ($a_{{\mathbf k}_{1}}$ and
$a_{{\mathbf k}_{1}^{\prime }}$) and the single mode on the second ring
($a_{{\mathbf k}_{2}}$) are joined to form a set of three variables
$\{a_{{\mathbf k}_{1}},a_{{\mathbf k}_{1}^{\prime }},a_{{\mathbf
k}_{2}}\}$, with the vectors satisfying ${\mathbf k}_{1}+{\mathbf
k}_{1}^{\prime }+{\mathbf k}_{2}=0$, their configuration specified by a
pair of parameters $\left( k_{1},k_{2}\right) $ up to a rotation.
Fig.~\ref{fig-rings} shows one example of such a configuration.
\begin{figure}
\centering
\epsfig{file=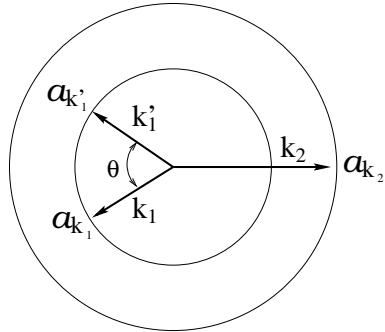, width=5cm}

\caption{A configuration of modes $ \{a_{{\mathbf k}_{1}},a_{{\mathbf
k}_{1}^{\prime }},a_{{\mathbf k}_{2}}\}$ with $ {\mathbf k}_{1}+{\mathbf
k}_{1}^{\prime }+{\mathbf k}_{2}=0$, where $ \left| {\mathbf k}_{1}\right|
=\left| {\mathbf k}_{1}^{\prime }\right| =k_{1}$ and $ \left| {\mathbf
k}_{2}\right| =k_{2}$ are the radii of two rings in Fourier
space.\label{fig-rings}}
\end{figure}

\item Equation (\ref{eq:chi111def}) with $x,y,z=a_{{\mathbf
k}_{1}},a_{{\mathbf k}_{1}^{\prime }},a_{{\mathbf k}_{2}}$ is employed to
calculate the estimator $\hat{\chi }_{111}({\mathbf k}_{1},{\mathbf
k}_{1}^{\prime },{\mathbf k}_{2})\equiv \hat{\chi }_{111}(k_{1},k_{2})$,
averaging over rotations. The estimator is then normalized to its standard
deviation calculated under the assumption of Gaussianity, i.e.~divided by
the square root of equation (\ref{eq:varchi21}). The resulting estimator is
\[
\hat{\bar{\chi }}_{111}(k_{1},k_{2})=\frac{\hat{\chi
}_{111}(k_{1},k_{2})}{\sqrt{\textrm{var}\left[ \hat{\chi }_{111}\left(
k_{1},k_{2}\right) \right] }}\]
\begin{equation}
\label{chi111-norm}
=\frac{\left[ N\left( k_{1},k_{2}\right) \right] ^{1/2}\hat{\chi
}_{111}(k_{1},k_{2})}{P(k_{1})P^{1/2}(k_{2})}.
\end{equation}
Here $N\left( k_{1},k_{2}\right) $ is the number of available samples.
\item If the underlying random field is Gaussian, then the estimators
$\hat{\bar{\chi }}_{111}(k_{1},k_{2})$ for all available $\left(
k_{1},k_{2}\right) $ should be independent Gaussian variables with mean
zero and variance one. Therefore, we further calculate the first few
cumulants of $\hat{\bar{\chi }}_{111}$ in the domain $(k_{1},k_{2})$. These
cumulants can be expressed as
\begin{eqnarray}
\textrm{mean}\left[ \hat{\bar{\chi }}_{111}\right] \equiv  &
\hat{\varepsilon }_{1} & \equiv \left\langle \hat{\bar{\chi
}}_{111}(k_{1},k_{2})\right\rangle ,\label{mu-chii} \\ \textrm{var}\left[
\hat{\bar{\chi }}_{111}\right] \equiv  & \hat{\varepsilon }_{2} & \equiv
\left\langle \left[ \hat{\bar{\chi }}_{111}(k_{1},k_{2})-\hat{\varepsilon
}_{1}\right] ^{2}\right\rangle ,\\ & \hat{\varepsilon }_{3} & \equiv
\left\langle \left[ \hat{\bar{\chi }}_{111}(k_{1},k_{2})-\hat{\varepsilon
}_{1}\right] ^{3}\right\rangle ,\\ & \hat{\varepsilon }_{4} & \equiv
\left\langle \left[ \hat{\bar{\chi }}_{111}(k_{1},k_{2})-\hat{\varepsilon
}_{1}\right] ^{4}\right\rangle -3\hat{\varepsilon }_{2}^{2},\\ &
\hat{\varepsilon }_{5} & \equiv \left\langle \left[ \hat{\bar{\chi
}}_{111}(k_{1},k_{2})-\hat{\varepsilon }_{1}\right] ^{5}\right\rangle
-10\hat{\varepsilon }_{2}\hat{\varepsilon }_{3}.
\end{eqnarray}
Here the ensemble averages are taken over all available pairs of $\left(
k_{1},k_{2}\right) $. We shall denote the number of samples in the domain
$(k_{1},k_{2})$ as $N_{\chi }$.
\item If the joint distribution of $\{a_{{\mathbf k}_{1}},a_{{\mathbf
k}_{1}^{\prime }},a_{{\mathbf k}_{2}}\}$ is Gaussian, then $\varepsilon
_{n}$ should have mean $0$, except $\varepsilon _{2}$ having mean $1$. They
also have variances
\begin{equation}
\textrm{var}\left[ \varepsilon _{n}\right] =\frac{n!}{N_{\chi }}.
\end{equation}
Therefore, if the estimators $\hat{\varepsilon }_{n}$ of the map depart
from their means much further than $\textrm{var}\left[ \varepsilon
_{n}\right] $, we can reject the hypothesis that the map is Gaussian. We
normalize the quantities $\hat{\varepsilon }_{n}$ to their variances:
\[
\hat{\omega }_{1}=\sqrt{N_{\chi }}\, \hat{\varepsilon }_{1},\; \;
\hat{\omega }_{2}=\sqrt{\frac{N_{\chi }}{2}}\left( \hat{\varepsilon
}_{2}-1\right) ,\]
\begin{equation}
\label{w_i}
\hat{\omega }_{n}=\sqrt{\frac{N_{\chi }}{n!}}\hat{\varepsilon }_{n},\:
n=3,4,5.
\end{equation}

\end{enumerate}
With this procedure, we evaluate the estimators $\hat{\omega }_{n}$ from
CMB maps under investigation. For a Gaussian map and sufficiently large
$N_{\chi }$, the estimators $\hat{\omega }_{n}$ are independent Gaussian
variables of mean zero and variance one. If some of the observed
$\hat{\omega }_{n}$ depart significantly from zero, we reject the
hypothesis that the map is Gaussian.

\subsection{The Gaussian background and instrumental noise}

\label{the-gaussian-background-and-instrument-noise}

Even if the main underlying mechanism for producing the CMB anisotropies
contributes a non-Gaussian compoment, the observed CMB may be close to
Gaussian due to the central limit theorem. We model this by adding a
Gaussian background to our simulated non-Gaussian map. This Gaussian
component is also expected in scenarios where both inflation and defects
are present. For simplificy, we assume instrumental noise to be Gaussian as
well. As a first step for testing the method, we compute the estimators
$\hat{\omega }_{n}$ for these two Gaussian components and their combination
and check numerically that the normalized cumulant estimators $\hat{\omega
}_{n}$ lie within the range $(-1,1)$ for these cases.
\begin{figure}
\centering
\epsfig{file=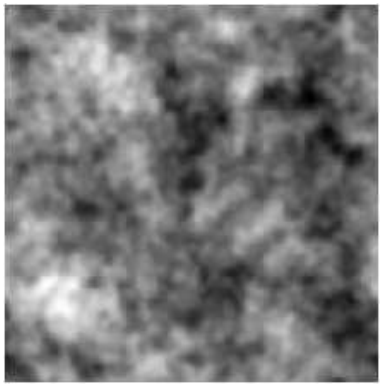, width=5cm}

\caption{Sample $ (5^{\circ })^{2}$ map of Gaussian background.
\label{fig:gau}}
\end{figure}
\begin{figure}
\centering
\epsfig{file=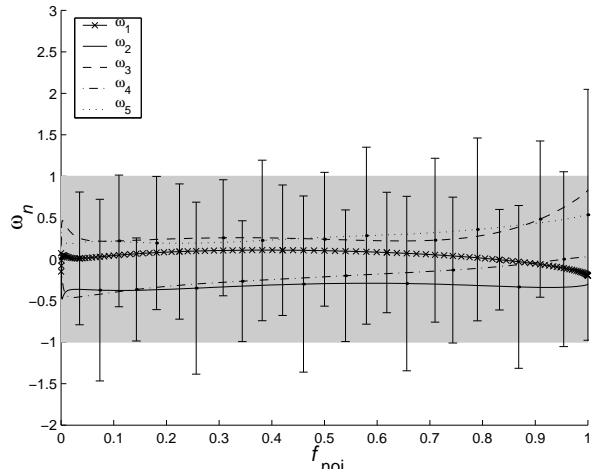, width=8cm}

\caption{Plot of the estimators $ \hat{\omega }_{n}$ of the $ (5^{\circ
})^{2}$ Gaussian map mixed with pixel noise as functions of the noise
fraction $ f_{\textrm{noi}}$. The shaded area is the one-sigma interval $
-1<\hat{\omega }_{n}<1$. \label{fig-wi-noi}}
\end{figure}

\begin{figure}
\centering
\epsfig{file=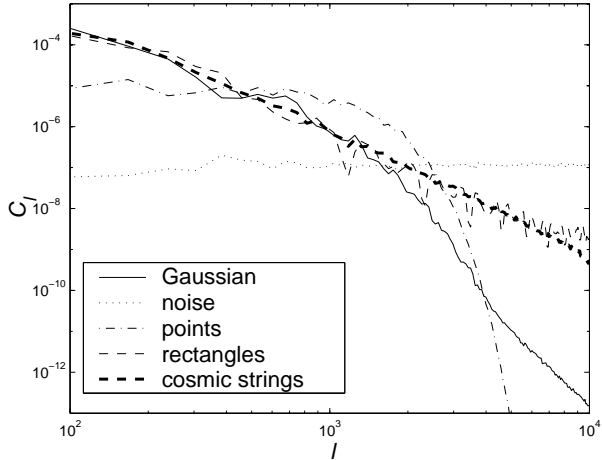, width=8cm}

\caption{Power spectra of various components presented in
Figs.~\ref{fig:gau}, \ref{fig:pnt}, \ref{fig:rect}, and \ref{fig:str}.
Their pixel standard deviations are all made equal to $ 1$ for comparison
purposes.  \label{fig-pow}}
\end{figure}

First, the Gaussian background is generated with the power spectrum
computed by CMBFAST \cite{CMBFAST} with sample parameters $\Omega
_{b}=0.05$, $\Omega _{\textrm{CDM}}=0.95$, $h=0.5$. We show one such map in
Fig.~\ref{fig:gau}. The instrumental noise is simulated as a Gaussian field
with a white-noise power spectrum. These two maps are mixed, with the noise
being a fraction
\begin{equation}
f_{\textrm{noi}}=\frac{\sigma ^{2}_{\textrm{noi}}}{\sigma
^{2}_{\textrm{G}}+\sigma ^{2}_{\textrm{noi}}},
\end{equation}
where $\sigma _{\textrm{G}}$ and $\sigma _{\textrm{noi}}$ denote
the RMS amplitudes of the Gaussian background and the white noise
respectively. We then compute the estimators $\hat{\omega }_{n}$ as defined
in Eq.~(\ref{w_i}), for the value of $f_{\textrm{noi}}$ going from zero to
unity. The results are plotted in Fig.~\ref{fig-wi-noi}. Each curve for an
estimator $\hat{\omega }_{n}$ is a mean of ten independent realizations
with error bars corresponding to the numerically obtained standard
deviation of that estimator. The same applies to later figures of this
type. For the range $0<f_{\textrm{noi}}<1$, we see that all $\hat{\omega
}_{n}$ lie within the one-sigma region $(-1,1)$ and the error bars are well
confined within the two-sigma region. This verifies the reliability of our
numerical procedure, as well as the theory about the use of $\hat{\omega
}_{n}$. For reference, we plot the power spectra of the Gaussian background
and the instrumental noise in Fig.~\ref{fig-pow}, as the solid and dotted
lines respectively.

We generated 10,000 realizations of Gaussian background and computed
$\hat{\omega }_{n}$ for each realization. This allowed us to check the
probability distributions of $\hat{\omega }_{n}$ numerically, and we find
that they are indeed Gaussian distributed with variances all equal to unity
for $n=1,...,5$. We have also numerically verified that this result is
independent of the underlying power spectrum of the Gaussian background, as
it should be on theoretical grounds. One of the advantages of employing
$\hat{\omega }_{n}$ for testing Gaussianity is that their theoretical
distributions are known, so that one does not need to implement Monte Carlo
simulations for the likelihood analysis, i.e.~to compute the ``equivalent
Gaussian realizations'' from the map being tested. In the likelihood
analyses below we shall simply shade the one-sigma area $-1<\hat{\omega
}_{n}<1$ in all relevant plots.

\subsection{Point sources}

One important challenge for all CMB data analysis is to deal with the
presence of point sources. This usually unwanted component obscures the
cosmological non-Gaussian signal in most CMB non-Gaussian tests. For this
reason it is important to see how this component contributes to our
estimators $\hat{\omega }_{n}$.

\begin{figure}
\centering
\epsfig{file=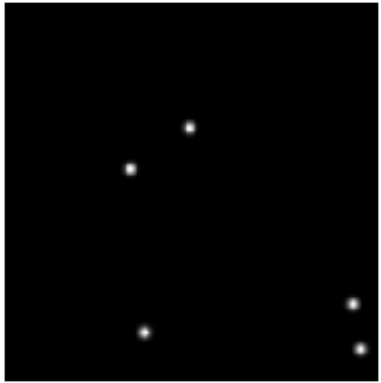, width=5cm}

\caption{A sample map of random point sources. \label{fig:pnt}}
\end{figure}

\begin{figure}
\centering
\epsfig{file=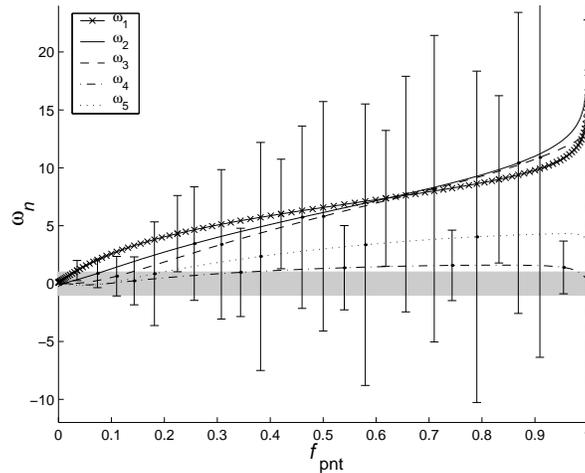, width=8cm}

\caption{Estimators $ \hat{\omega }_{n}$ of the map of random point
sources mixed with the Gaussian background, as functions of the fraction of
power in the point sources $ f_{\textrm{pnt}}$. The shaded
area is $ -1<\hat{\omega }_{n}<1$.  \label{fig-wi-pnt}}
\end{figure}

We first generate five point sources, each having the power spectrum
\begin{equation}
\label{eqn-pnt}
C_{l}\propto \exp \left[ -\left( rRk\right) ^{2}\right] ,
\end{equation}
where $R$ is the angular size of the field. Thus $r$ indicates the
size of a point as a fraction of the size of the field, and here we use
$r=0.01$. A sample map is shown in Fig.~\ref{fig:pnt}. The map of point
sources is superimposed onto the same Gaussian background as in the
previous subsection, with a fraction of the point sources
\begin{equation}
f_{\textrm{pnt}}=\frac{\sigma ^{2}_{\textrm{pnt}}}{\sigma
^{2}_{\textrm{G}}+\sigma ^{2}_{\textrm{pnt}}},
\end{equation}
where $\sigma _{\textrm{pnt}}$ is the RMS amplitude of the point signal.
We then compute the estimators $\hat{\omega }_{n}(f_{\textrm{pnt}})$ as
in the previous case. The results are plotted in Fig.~\ref{fig-wi-pnt}.

As one can see, when the fraction of the point signal is weak enough
($f_{\textrm{pnt}}\lesssim 0.02$), its non-Gaussianity does not show up.
Once this signal becomes stronger ($f_{\textrm{pnt}}\gtrsim 0.02$),
$\hat{\omega }_{1}$ starts departing from the one-sigma region while the
other $\hat{\omega }_{n}$ follow at larger $f_{\textrm{pnt}}$. With
$f_{\textrm{pnt}}\gtrsim 0.1$, one can use $\hat{\omega }_{1}$ to reject
the hypothesis of Gaussianity at a confidence level of more than $99\%$,
since the solid line with crosses goes outside the three-sigma range
$(-3,3)$.

We plotted the power spectrum of the point sources as the dot-dashed line
in Fig.~\ref{fig-pow}. By comparing this line to the solid line (the power
spectrum of the Gaussian background), we find that for
$f_{\textrm{pnt}}\lesssim 0.02$, the point signal never comes to dominate
on any scale and so unsurprisingly passes the Gaussianity test using
$\hat{\omega }_{n}$. Once $f_{\textrm{pnt}}\gtrsim 0.02$, the power
spectrum of the point signal starts to dominate at $l\approx 2000$, at
which stage it fails the test. Thus, we conclude that our estimators
$\hat{\omega }_{n}$ are sensitive to point sources. For this method to
succeed in real CMB data, therefore, it will be necessary to first remove
the point sources using any of the available methods (see
e.g.~\cite{remove-point-cmb}), and then compute the estimators $\hat{\omega
}_{n}$. In the following analyses, we shall ignore the point sources by
assuming that they have been removed beforehand.

\subsection{Test on filled rectangles}

\label{test-on-filled-rectangles}

\begin{figure}
\centering
\epsfig{file=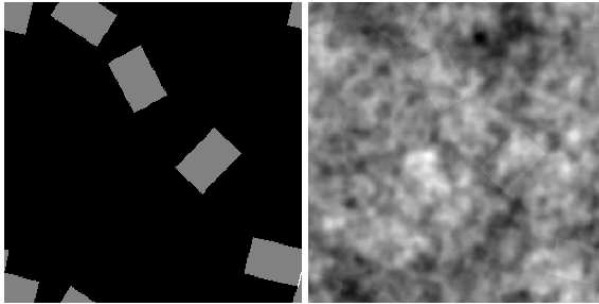, width=8cm}

\caption{A sample map of filled rectangles (left) and the same map mixed
with a Gaussian background with a fraction of power in the rectangles $
f_{\textrm{rect}}=0.0005$ (right).\label{fig:rect}}
\end{figure}

\begin{figure}
\centering
\epsfig{file=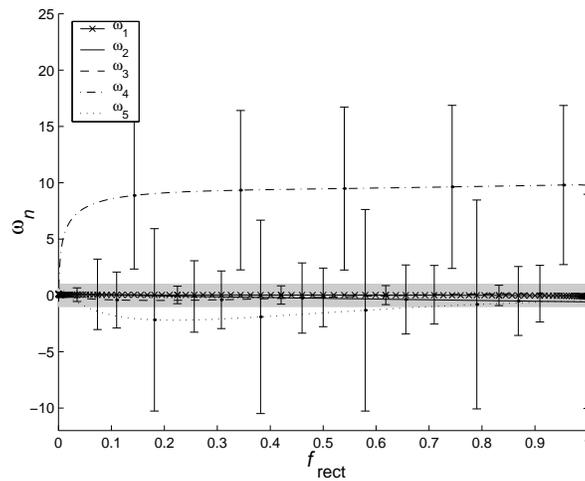, width=8cm}

\caption{Plot of the estimators $ \hat{\omega }_{n}$ of the mixed
map of rectangles with Gaussian background as a function of $
f_{\textrm{rect}}$. The shaded area is $ -1<\hat{\omega }_{n}<1$. 
\label{fig-wi-rect}}
\end{figure}

To test the sensitivity of our method to certain types of non-Gaussianity,
we first try a simple model using filled rectangles. In a $(5^{\circ
})^{2}$ field with a resolution of $256^{2}$ pixels, we generate five
filled rectangles, each having a size of $32\times 48$ grid spacings. These
rectangles are shown in Fig.~\ref{fig:rect}. This map is then mixed with
the Gaussian background used in Fig.~\ref{fig-wi-noi}, with the fraction of
rectangles $f_{\textrm{rect}}$ defined similarly to the fraction of point
sources $f_{\textrm{pnt}}$. We then computed the estimators $\hat{\omega
}_{n}(f_{\textrm{rect}})$ as before, and the results are plotted in
Fig.~\ref{fig-wi-rect}.

We find that one can use $\hat{\omega }_{4}$ to reject the hypothesis of
Gaussianity at more than a 95\% confidence level when
$f_{\textrm{rect}}\gtrsim 0.001$. This means that non-Gaussian features of
the rectangles show up once their RMS amplitude is larger than a few
percent of the total amplitude. Referring to the power spectrum of the
rectangles in Fig.~\ref{fig-pow} (the dashed line), we see that once
$f_{\textrm{rect}}\lesssim 0.001$, the power of the rectangles never comes
to dominate when compared with the power of the Gaussian background (the
solid line). This explains why non-Gaussian features of the rectangles are
not visible in $\hat{\omega }_{n}$ when $f_{\textrm{rect}}\lesssim 0.001$.

\begin{figure}
\centering
\epsfig{file=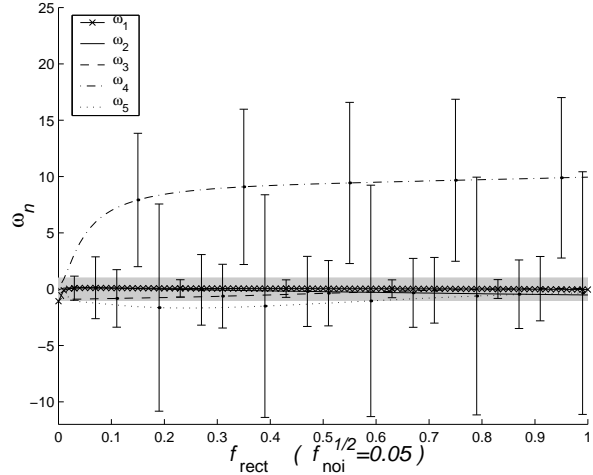, width=8cm}

\caption{Same as Fig.~\ref{fig-wi-rect} but mixed with noise with the
fraction of power $ f_{\textrm{noi}}=0.05^{2}$. \label{fig-wi-rect-noi}}
\end{figure}

We also test the sensitivity of our method to the rectangles in the
presence of noise. In this case we mix three components: the rectangles,
the Gaussian background, and the white noise. The last two components are
exactly the same as the two we used in
Section~\ref{the-gaussian-background-and-instrument-noise}. The noise
fraction is fixed as $f_{\textrm{noi}}=0.05^{2}$, so the rectangles have a
fraction
\begin{equation}
f_{\textrm{rect}}=\frac{\sigma ^{2}_{\textrm{rect}}}{\sigma
^{2}_{\textrm{rect}}+\sigma ^{2}_{\textrm{G}}+\sigma
^{2}_{\textrm{noi}}}=1-0.05^{2}-f_{\textrm{G}}\, .
\end{equation}
The $\hat{\omega }_{n}$ are computed as a function of $f_{\textrm{rect}}$,
and the results are presented in Fig.~\ref{fig-wi-rect-noi}.

This time we see that the non-Gaussian feature of the rectangles does not
show up until a higher $f_{\textrm{rect}}$. For example, one can use
$\hat{\omega }_{4}$ to reject the hypothesis of Gaussianity at more then
95\% confidence level only when $f_{\textrm{rect}}\gtrsim 0.02$. Referring
to the power spectra of the rectangles, the Gaussian background, and the
noise in Fig.~\ref{fig-pow}, one finds that when $f_{\textrm{rect}}\lesssim
0.02$ with $f_{\textrm{noi}}=0.05^{2}$, the power of the rectangles is
dominated by the power of the Gaussian components (the Gaussian background
and the noise); but when $f_{\textrm{rect}}\gtrsim 0.02$, the power in
rectangles starts to dominate at $l\approx 4000$. This verifies again that
for $\hat{\omega }_{n}$ to succeed in detecting a non-Gaussian signal, the
power of the the non-Gaussian component needs to dominate within at least a
certain range of the accessible $l$. Thus we see that noise may reduce the
sensitivity of our method for detecting non-Gaussian signals, depending on
whether the noise dominates the non-Gaussian signal on scales where it
originally dominated.

\subsection{Detecting cosmic strings}

\begin{figure}
\centering
\epsfig{file=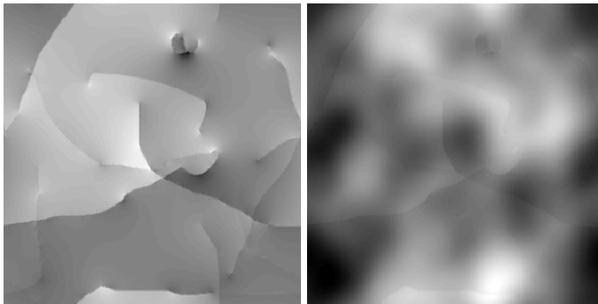, width=8cm}

\caption{Sample $ (1^{\circ })^{2}$ maps of the string-induced ISW
effect (left) and the same with added Gaussian background (right) where the
fraction of power in the string component is $
f_{\textrm{str}}=0.0001$.\label{fig:str}}
\end{figure}

\begin{figure}
\centering
\epsfig{file=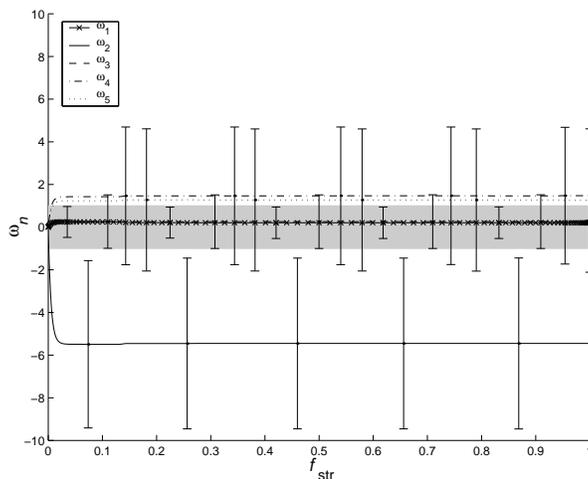, width=8cm}

\caption{Estimators $ \hat{\omega }_{n}$ of the map of strings mixed
with a Gaussian background as functions of $ f_{\textrm{str}}$.
The shaded area is $ -1<\hat{\omega }_{n}<1$.  \label{fig-wi-str}}
\end{figure}

Finally, we test our method against the string-induced CMB map. We use a
toy model of Ref.~\cite{Proty} to simulate the string-induced integrated
Sachs-Wolfe (ISW) effect. This model incorporates most main features of
cosmic strings, such as the scaling and self-avoiding properties, as well
as their wiggliness. One realization of the resulting CMB is shown in
Fig.~\ref{fig:str}. The dynamic range is $40$ in conformal time starting
from last scattering, and the angular size is $(1^{\circ })^{2}$ with a
resolution of $256^{2}$. A Gaussian background is simulated as before.
These two maps are then linearly summed, with a string fraction
$f_{\textrm{str}}$ defined as above. The estimators $\hat{\omega }_{n}$ are
calculated as functions of $f_{\textrm{str}}$, and the results are shown in
Fig.~\ref{fig-wi-str}.

From the solid line in the figure ($\hat{\omega }_{2}$), one can reject
the hypothesis of Gaussianity at more than a 95\% confidence level when
$f_{\textrm{str}}\gtrsim 0.001$. The power spectrum of the Gaussian
background is shown as the solid line in Fig.~\ref{fig-pow}, while that of
the string-induced ISW effect is shown with the thick dashed line in the
same plot. Again, if we compare these two lines, we find that when
$f_{\textrm{str}}\lesssim 0.001$, the string-induced perturbations are
dominated by the Gaussian background on all scales; when
$f_{\textrm{str}}\gtrsim 0.001$, the string-induced perturbations start to
dominate at $l\sim 10^{4}$. This result is consistent with our previous
argument about the sensitivity of the estimators $\hat{\omega }_{n}$.

\begin{figure}
\centering
\epsfig{file=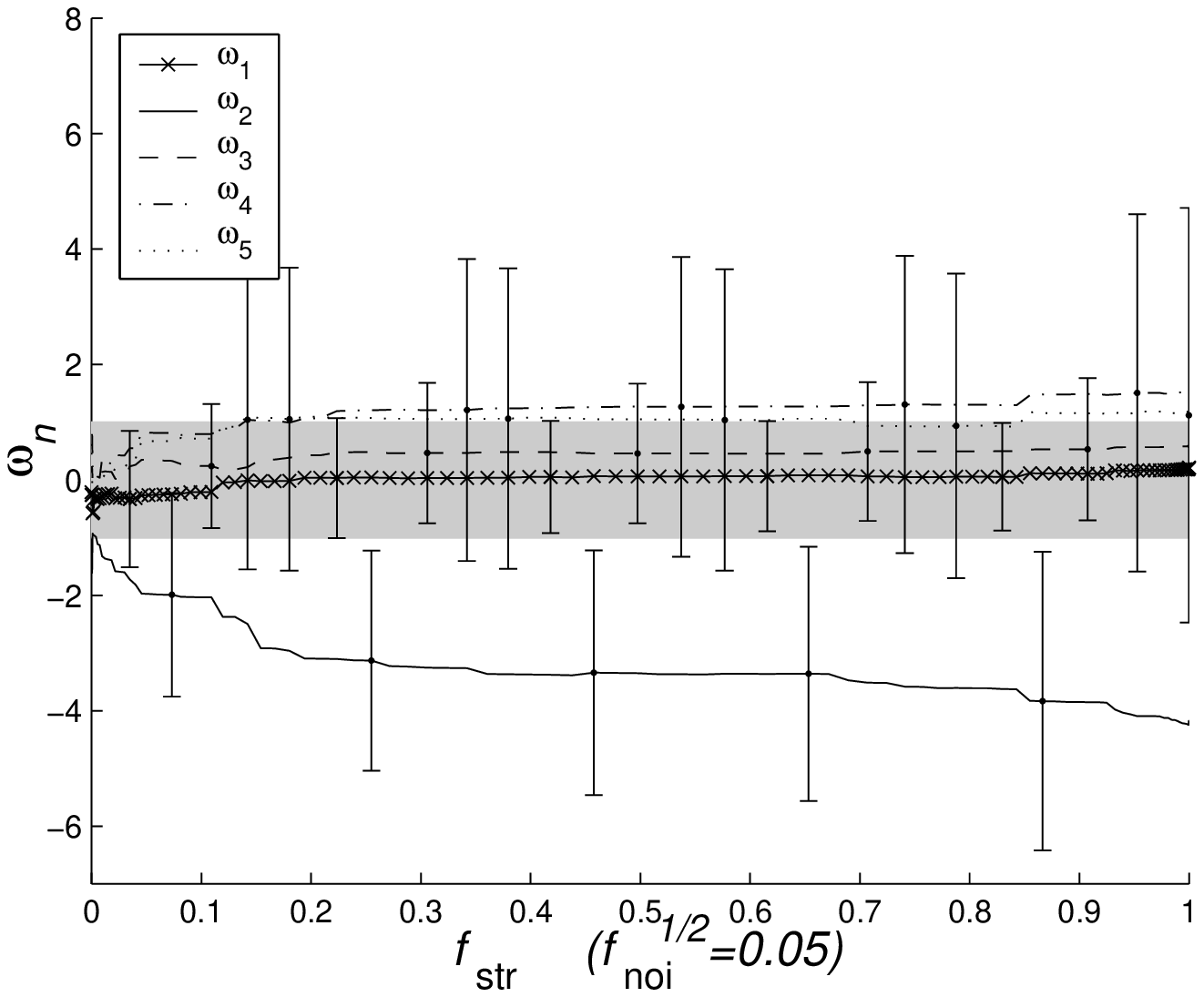, width=8cm}

\caption{Same as Fig.~\ref{fig-wi-str} but for maps mixed with noise, with
a noise fraction $ f_{\textrm{noi}}=0.05^{2}$.  \label{fig-wi-str-noi}}
\end{figure}

In the presence of instrumental noise, this sensitivity will be reduced if
the noise dominates the string-induced perturbations on scales where they
originally dominated. In other words, the presence of the noise will
contribute extra power to the Gaussian component of the underlying map, so
as to raise the threshold in power beyond which the non-Gaussian signal may
dominate. This argument is again verified in Fig.~\ref{fig-wi-str-noi},
where we mix three components: the string-induced ISW effect, the Gaussian
background, and the white noise whose strength is $5\%$ of the total RMS
amplitude. This gives
\begin{equation}
f_{\textrm{str}}=\frac{\sigma ^{2}_{\textrm{str}}}{\sigma
^{2}_{\textrm{str}}+\sigma ^{2}_{\textrm{G}}+\sigma
^{2}_{\textrm{noi}}}=1-0.05^{2}-f_{\textrm{G}}\, .
\end{equation}
As one can see, the hypothesis of Gaussianity is rejected at more than a
95\%
confidence level when $f_{\textrm{str}}\gtrsim 0.05$, using $\hat{\omega
}_{2}$. This threshold $f_{\textrm{str}}\approx 0.05$ can be again verified
by comparing the thick dashed, solid, and dotted lines in
Fig.~\ref{fig-pow}.

By comparing results for strings and rectangles (see
section~\ref{test-on-filled-rectangles}), we find that they have similar
thresholds for rejecting the hypothesis of Gaussianity. This is because
strings and rectangles have almost identical power spectra and they
dominate the Gaussian component above similar thresholds and at similar
scales.

By comparing Figs.~\ref{fig-wi-str} and \ref{fig-wi-str-noi}, with
Figs.~\ref{fig-wi-pnt}, \ref{fig-wi-rect} and \ref{fig-wi-rect-noi}, we
find that only string-induced perturbations result in a significantly
negative value of the estimator $\hat{\omega }_{2}$, while all other
non-Gaussian components we tried produce only positive $\hat{\omega }_{n}$
outside the 95\% confidence region (the two-sigma region). This provides a
potential discriminator in distinguishing string-induced features from
other non-Gaussian signals.

\section{Conclusion}

\label{cmb-ng2-conclusion}

In this paper, we studied the properties of normalized inter-scale
cumulants of Fourier modes for homogeneous random fields and proposed a new
set of statistics to test a small-scale CMB map for Gaussianity. We showed
that higher-order cumulant estimators of Fourier components of Gaussian
fields are themselves nearly Gaussian distributed variables with zero
expectation value and known variances. Therefore, a test of Gaussianity of
these cumulant estimators computed from a given map constitutes a test of
Gaussianity of the map. This method is quite general and can be employed on
cumulants of any order. However, since the usefulness of higher-order
cumulants quickly drops with the order, the bispectrum components and the
4-th order correlators are the most promising candidates.

As an application, the statistical estimators that we denoted by
$\hat{\omega }_{n}$ were constructed from all available bispectrum
components. We prepared simulated maps made up of random point sources,
random rectangles, and simulated cosmic string networks, superimposed on
Gaussian background and pixel noise. By developing an analytic model of
fields containing random superimposed shapes, we showed that inter-scale
cumulants of all orders are generically present in such fields, and that
noise and Gaussian background limits the sensitivity of the cumulant
estimators on scales where its power spectrum is significant compared to
the power spectrum of the non-Gaussian component.

We numerically verified that the discriminators $\hat{\omega }_{n}$ are
sensitive to non-Gaussian signal of simulated maps, as well as to point
sources. To apply these statistics to detect certain types of non-Gaussian
signals such as those resulting from defects, one needs therefore to remove
point sources on scales where they dominate the non-Gaussian components one
would like to detect. The instrumental noise was also shown to be capable
of reducing the sensitivity of our new method, due to its extra
contribution to the Gaussian component on small scales. Nevertheless, as we
theoretically showed and numerically verified, our new statistics are
capable of detecting non-Gaussian component as long as its power spectrum
dominates within some range of accessible scales. We also found that the
string-induced ISW effect, unlike other non-Gaussian models we tested, such
as point sources, induces a significantly negative value of $\hat{\omega
}_{2}$. This is a potential discriminator of string-induced perturbations.
We intend to apply this method to characterize the non-Gaussian features
resulting from more realistic defect models. Even if defects are
unimportant as the underlying mechanism for structure formation, we could
still apply these methods to detect their existence, especially when
high-accuracy and high-resolution observations become available in the near
future.

\section{Acknowledgments}

The authors are grateful to Andrew Liddle, Paul Shellard, and Neil Turok
for useful comments and to Alan Heavens, Jo\~ao Magueijo, and Dmitry
Pogosian for helpful discussions. A considerable part of this work was
completed at DAMTP, University of Cambridge, Cambridge CB3 9EW, U.K.
S.~W.~is supported by DOE and J.-H.~P.~W.~is funded by NSF KDI Grant
(9872979) and NASA LTSA Grant (NAG5-6552).C.

\appendix

\section{Distribution of cumulant estimators for independent Gaussian
samples }

\label{app:dc}To test whether a random variable $x$ is Gaussian
distributed, one can estimate higher-order cumulants $\chi _{n}$, $n>2$ of
that variable and check that they vanish within their statistical variance.
Given a number of realizations of $x$, one can compute moment estimators
$\hat{\mu }_{n}$ and cumulant estimators $\hat{\chi }_{n}$ for that
purpose. However, obtaining the likelihood that the data $x_{i}$ satisfies
the hypothesis requires knowledge of the distribution of the cumulant
estimators $\hat{\chi }_{n}$ for an underlying hypothetical Gaussian
ensemble of realizations of $x$. If the cumulant estimators $\hat{\chi
}_{n}$ were themselves Gaussian distributed, one would only need to know
their ensemble mean values $\left\langle \hat{\chi }_{n}\right\rangle $ and
covariances $\left\langle \hat{\chi }_{m}\hat{\chi }_{n}\right\rangle $.

In this Appendix we give a self-contained derivation for the covariances of
the cumulant estimators $\hat{\chi }_{m}$ computed for a set of $N$
independent samples of one (Gaussian) random variable $x$ with mean $\mu $
and standard deviation $\sigma $. We show below that to leading order in
$N^{-1}$, the cumulant estimators are biased by
\begin{eqnarray}
\left\langle \hat{\chi }_{2n}\right\rangle  & = & \chi
_{2n}-N^{-1}\frac{\left( 2n\right) !}{2n!}\sigma ^{2n}+O\left(
N^{-2}\right) ,\label{chi1anse} \\ \left\langle \hat{\chi
}_{2n+1}\right\rangle  & = & \chi _{2n+1}+O\left( N^{-2}\right)
,\label{chi1anso}
\end{eqnarray}
and are statistically independent,
\begin{equation}
\label{chi2ans}
\left\langle \hat{\chi }_{m}\hat{\chi }_{n}\right\rangle -\left\langle
\hat{\chi }_{m}\right\rangle \left\langle \hat{\chi }_{n}\right\rangle
=N^{-1}\sigma ^{2n}n!\delta _{mn}+O\left( N^{-2}\right) .
\end{equation}
Here $\chi _{1}=\mu $, $\chi _{2}=\sigma ^{2}$ and $\chi _{n}=0$
for $n\geq 3$. We shall also show that the higher-order correlations
between $\chi _{n}$ such as $\left\langle \hat{\chi }_{l}\hat{\chi
}_{m}\hat{\chi }_{n}\right\rangle $ are absent up to terms of order
$O\left( N^{-2}\right) $ and therefore for large $N$ these estimators can
be treated as approximately Gaussian distributed, independent random
variables with known means and variances. This is immediately relevant to
inter-scale cumulants of Fourier space amplitudes, and by using our
formalism we show that these are also independent.

Consider a set of $N$ independent realizations $x_{i}$, $i=1,...,N$
of the Gaussian random variable $x$. This set satisfies
\begin{equation}
\label{gauss-di}
\left\langle x_{i}\right\rangle =\mu ,\quad \left\langle
x_{i}x_{j}\right\rangle =\mu ^{2}+\sigma ^{2}\delta _{ij}.
\end{equation}
The problem at hand is to characterize the estimators for the cumulants
$\chi _{n}$ of the distribution of $x$. First we take the unbiased
estimators $\hat{\mu }_{n}$ of the moments of $x$, which are
\begin{equation}
\hat{\mu }_{n}\equiv N^{-1}\sum _{i}x_{i}^{n}.
\end{equation}
The cumulant estimators $\hat{\chi }_{n}$ may be defined through these
moment estimators by the usual formulae relating cumulants and moments,
e.g. for the second and third cumulants
\begin{equation}
\label{cumu2}
\hat{\chi }_{2}=\hat{\mu }_{2}-\hat{\mu }_{1}^{2},
\end{equation}
\begin{equation}
\hat{\chi }_{3}=\hat{\mu }_{3}-3\hat{\mu }_{2}\hat{\mu }_{1}+2\hat{\mu
}_{1}^{3},
\end{equation}
and so on (these standard relations follow from Eqs.
(\ref{mom-gen})-(\ref{cumu-gen}) below). Defined in this manner, the
cumulant estimators $\hat{\chi }_{n}$ are however not unbiased, because for
instance $\hat{\mu }_{1}^{2}$ in Eq. (\ref{cumu2}) is a biased estimator
for the square of the first moment $\mu _{1}^{2}$:
\begin{equation}
\label{mu1sq}
\left\langle \hat{\mu }_{1}^{2}\right\rangle =\left\langle N^{-2}\sum
_{ij}x_{i}x_{j}\right\rangle =\mu _{1}^{2}+N^{-1}\sigma ^{2}.
\end{equation}
We shall show now that the resulting bias in $\hat{\chi }_{n}$ is always
of order $N^{-1}$ or smaller.

We need to consider this bias in more detail since the expectation values
of $\chi _{n}$ are zero for $n>2$ and also because terms of order $O\left(
N^{-1}\right) $ dominate the variance of the estimators $\hat{\chi }_{n}$.
Terms of order $O\left( N^{-1}\right) $ arise in expressions such as
Eq.~(\ref{cumu2}) from products of moment estimators. By analogy with Eq.
(\ref{mu1sq}) one obtains
\begin{equation}
\label{muamub}
\left\langle \hat{\mu }_{a}\hat{\mu }_{b}\right\rangle =\mu _{a}\mu
_{b}+N^{-1}\left( \mu _{a+b}-\mu _{a}\mu _{b}\right) .
\end{equation}
Products of more than two moment estimators will also contain terms of
higher order than $N^{-1}$ but here we are only interested in the leading
terms. For products of three moments, we obtain
\begin{equation}
\left\langle \hat{\mu }_{a}\hat{\mu }_{b}\hat{\mu }_{c}\right\rangle =\mu
_{a}\mu _{b}\mu _{c}+N^{-1}\left( \mu _{a+b}\mu _{c}+\mu _{b+c}\mu _{a}+\mu
_{c+a}\mu _{b}-3\mu _{a}\mu _{b}\mu _{c}\right) +O\left( N^{-2}\right) ,
\end{equation}
and it is straightforward to generalize to products of $k$ estimators,
\begin{equation}
\label{nmoments}
\left\langle \hat{\mu }_{a_{1}}...\hat{\mu }_{a_{k}}\right\rangle =\mu
_{a_{1}}...\mu _{a_{k}}+N^{-1}\sum _{1\leq i<j\leq k}\left[ \mu
_{a_{1}}...\mu _{a_{i}+a_{j}}...\mu _{a_{k}}-\mu _{a_{1}}...\mu
_{a_{k}}\right] +O\left( N^{-2}\right) .
\end{equation}
The sum above is performed over all pairs of moments entering the original
expression.

Now we need to find out where and how such terms appear in expressions for
cumulant estimators. The relation of cumulants $\chi _{n}$ to moments $\mu
_{n}$ is most easily understood by means of the generating function of
moments (also called the characteristic function) $Z\left[ p\right] $ which
satisfies
\begin{equation}
\label{mom-gen}
\mu _{n}=\frac{1}{Z\left[ 0\right] }\left( i\frac{\partial }{\partial
p}\right) ^{n}Z\left[ p\right] _{p=0}.
\end{equation}
Because of the normalization $Z\left[ 0\right] ^{-1}$ the generating
function $Z\left[ p\right] $ is defined up to an irrelevant constant
factor. For instance, the Gaussian distribution from Eq.~(\ref{gauss-di})
has the generating function
\begin{equation}
\label{gaussz}
Z\left[ p\right] =\exp \left[ -\frac{p^{2}}{2}\sigma ^{2}-i\mu p\right] .
\end{equation}
The cumulants can be expressed (or, equivalently, defined) as
\begin{equation}
\label{cumu-gen}
\chi _{n}=\left( i\frac{\partial }{\partial p}\right) ^{n}\ln Z\left[
p\right] _{p=0}
\end{equation}
(note that the constant factor in the definition of $Z\left[ p\right] $
drops out). The $n$-th derivative of $\ln Z\left[ p\right] $ in Eq.
(\ref{cumu-gen}) will contain terms such as
\begin{equation}
\label{derivterms}
\frac{1}{Z\left[ p\right] }\frac{\partial ^{l}Z\left[ p\right] }{\partial
p^{l}}...\frac{1}{Z\left[ p\right] }\frac{\partial ^{m}Z\left[ p\right]
}{\partial p^{m}}
\end{equation}
which give rise to products of moments $\mu _{l}...\mu _{m}$. We would
like to replace the moments $\mu _{n}$ by their estimators $\hat{\mu }_{n}$
and then to combine them pairwise to obtain the $O\left( N^{-1}\right) $
terms as shown in Eq.~(\ref{nmoments}).

We notice that if we take an expression such as $\left( 1+N^{-1/2}\mu
_{1}\right) ...\left( 1+N^{-1/2}\mu _{k}\right) $ and expand it in
$N^{-1/2}$, then the term of order $O\left( N^{-1}\right) $ will be equal
to the sum over all pairs of the product $\mu _{i}\mu _{j}$. This is almost
what is required by Eq. (\ref{nmoments}), except that we also need to
somehow convert $\mu _{i}\mu _{j}$ into $\mu _{i+j}$. This would happen if
$\mu _{n}$ behaved like a differential operator $\left( \partial /\partial
t\right) ^{n}$ acting on functions of a dummy variable $t$. The required
technical trick is implemented as follows. We first replace the moments
$\mu _{n}$ in Eq. (\ref{cumu-gen}) by the differential operator $\left( \mu
_{n}+N^{-1/2}\left[ i\partial /\partial t\right] ^{n}\right) $. This is
done by changing the generating function of moments $Z\left[ p\right] $ to
\begin{equation}
\label{newzp}
\tilde{Z}\left[ p;t\right] \equiv \sum _{n=0}^{\infty }\frac{\left(
-ip\right) ^{n}}{n!}\left( \mu _{n}+N^{-1/2}i^{n}\frac{\partial
^{n}}{\partial t^{n}}\right) =Z\left[ p\right] +N^{-1/2}\exp \left[
p\partial _{t}\right] .
\end{equation}
Then we introduce another copy of the old generating function of moments,
$Z\left[ t\right] $, on which this new generating function $\tilde{Z}\left[
p;t\right] $ will act with $\partial _{t}$ at the end of the calculation,
so that the derivatives $\left[ i\partial _{t}\right] ^{n}$ will at the end
be replaced again by the moments $\mu _{n}$. We now claim that the
expectation value of cumulant estimators $\hat{\chi }_{n}$ and of their
products can be obtained up to and including terms of order $O\left(
N^{-1}\right) $ from the modified generating function $\tilde{Z}\left[
p;t\right] $ as
\begin{equation}
\label{avchi1}
\left\langle \hat{\chi }_{n}\right\rangle =\left[ i\partial _{p}\right]
^{n}\ln \tilde{Z}\left[ p;t\right] Z\left[ t\right] _{p=t=0}+O\left(
N^{-3/2}\right)
\end{equation}
and
\begin{equation}
\label{avchi2}
\left\langle \hat{\chi }_{m}\hat{\chi }_{n}\right\rangle =\left( \left[
i\partial _{p^{\prime }}\right] ^{m}\ln \tilde{Z}\left[ p^{\prime
};t\right] \right) \left( \left[ i\partial _{p}\right] ^{n}\ln
\tilde{Z}\left[ p;t\right] \right) Z\left[ t\right] _{p=p^{\prime
}=t=0}+O\left( N^{-3/2}\right) .
\end{equation}
Here the parameters $p$, $p^{\prime }$ and $t$ are set to $0$
only after all derivatives have been computed. Expectation values of
products of three or more cumulant estimators $\hat{\chi }_{n}$ are be
obtained in the same manner,

\begin{equation}
\left\langle \hat{\chi }_{n_{1}}...\hat{\chi }_{n_{k}}\right\rangle =\left(
\left[ i\partial _{p_{1}}\right] ^{n_{1}}\ln \tilde{Z}\left[ p_{1};t\right]
\right) ...\left( \left[ i\partial _{p_{k}}\right] ^{n_{k}}\ln
\tilde{Z}\left[ p_{k};t\right] \right) Z\left[ t\right]
_{p_{j}=t=0}+O\left( N^{-3/2}\right) .
\end{equation}

We shall first show that Eqs. (\ref{avchi1})-(\ref{avchi2}) actually give
the desired $O\left( N^{-1}\right) $ terms and then proceed to compute
these terms.

Consider Eq.~(\ref{avchi1}): after evaluating all derivatives in $p$ we
would obtain many terms of the form of Eq.~(\ref{derivterms}) but with the
modified generating function $\tilde{Z}\left[ p;t\right] $ acting (with
the derivatives in $t$) on $Z\left[ t\right] $,
\begin{equation}
\label{ntterm}
\left( \frac{1}{\tilde{Z}\left[ p;t\right] }\frac{\partial
^{l_{1}}\tilde{Z}\left[ p;t\right] }{\partial
p^{l_{1}}}...\frac{1}{\tilde{Z}\left[ p;t\right] }\frac{\partial
^{l_{n}}\tilde{Z}\left[ p;t\right] }{\partial p^{l_{n}}}\right)
_{p=0}Z\left[ t\right] _{t=0}.
\end{equation}
Take one such term, substitute $p=0$ and expand in $N^{-1/2}$:
\[
\left( \frac{1}{\tilde{Z}\left[ p;t\right] }\frac{\partial
^{l_{1}}\tilde{Z}\left[ p;t\right] }{\partial
p^{l_{1}}}...\frac{1}{\tilde{Z}\left[ p;t\right] }\frac{\partial
^{l_{n}}\tilde{Z}\left[ p;t\right] }{\partial p^{l_{n}}}\right)
_{p=0}Z\left[ t\right] _{t=0}\]

\[
=\frac{\mu _{l_{1}}+N^{-1/2}\left( i\partial _{t}\right)
^{l_{1}}}{1+N^{-1/2}}...\frac{\mu _{l_{n}}+N^{-1/2}\left( i\partial
_{t}\right) ^{l_{n}}}{1+N^{-1/2}}Z\left[ t\right] _{t=0}\]

\[
=\mu _{l_{1}}...\mu _{l_{n}}+N^{-1/2}\left( -n\mu _{l_{1}}...\mu
_{l_{n}}+\sum _{k=1}^{n}\mu _{l_{1}}...\left( i\partial _{t}\right)
^{l_{k}}...\mu _{l_{n}}\right) Z\left[ t\right] _{t=0}\]
\begin{equation}
\label{n1term}
+N^{-1}\left( -\frac{n\left( n-1\right) }{2}\mu _{l_{1}}...\mu
_{l_{n}}+\sum _{1\leq i<j\leq n}\mu _{l_{1}}...\left( i\partial _{t}\right)
^{l_{i+j}}...\mu _{l_{n}}\right) Z\left[ t\right] _{t=0}+O\left(
N^{-3/2}\right) .
\end{equation}
Note that after taking the derivatives in $t$ the $O\left( 1\right) $
and $O\left( N^{-1}\right) $ terms in Eq.~(\ref{n1term}) are exactly the
ones in Eq.~(\ref{nmoments}), while the $O\left( N^{-1/2}\right) $ terms
cancel. Therefore each term of the form of Eq.~(\ref{ntterm}) yields the
desired combination of moments for Eq.~(\ref{avchi1}). Note that although
the modified generating function $\tilde{Z}\left[ p;t\right] $ gives the
correct result in the $O\left( N^{-1}\right) $ terms, its higher-order
expansion terms are not useful.

The same argument can be shown to hold for Eq.~(\ref{avchi2}) and generally
for analogous expressions for averages of products of several $\hat{\chi
}_{n}$. We only need to take all derivatives in the parameters $p$,
$p^{\prime }$, ... prior to taking the derivative in $t$ in those
expansions.

Our strategy to obtain the expectation values from
Eqs.~(\ref{avchi1})-(\ref{avchi2}) will be to first expand in $N^{-1/2}$ up
to $O\left( N^{-1}\right) $, then substitute $p=0$ in the correct order,
and then evaluate the derivatives in $t$. Since we know that the $O\left(
1\right) $ terms give the unbiased result and that terms of order $O\left(
N^{-1/2}\right) $ vanish (both of these statements are straightforwardly
checked by a similar calculation), while the $O\left( N^{-3/2}\right) $ and
higher-order terms are not useful for us, we only concentrate on the
$O\left( N^{-1}\right) $ terms. Expanding the logarithm in
Eq.~(\ref{newzp}),
\begin{equation}
\label{logzte}
\ln \tilde{Z}\left[ p;t\right] =\ln Z\left[ p\right] +N^{-1/2}\frac{\exp
\left[ p\partial _{t}\right] }{Z\left[ p\right]
}-\frac{N^{-1}}{2}\frac{\exp \left[ 2p\partial _{t}\right] }{Z\left[
p\right] ^{2}}+O\left( N^{-3/2}\right) ,
\end{equation}
we obtain the expression
\begin{equation}
\label{lnzoneterm}
\ln \tilde{Z}\left[ p;t\right] Z\left[ t\right] =\chi _{n}Z\left[ t\right]
+N^{-1/2}\frac{Z\left[ p+t\right] }{Z\left[ p\right]
}-\frac{N^{-1}}{2}\frac{Z\left[ 2p+t\right] }{Z\left[ p\right]
^{2}}+O\left( N^{-3/2}\right) ,
\end{equation}
which will give the answer to Eq.~(\ref{avchi1}) if we take the derivative
in $p$ and substitute $p=t=0$. The result is
\begin{equation}
\label{chi1genans}
\left\langle \hat{\chi }_{n}\right\rangle =\chi _{n}-\frac{N^{-1}}{2}\left[
i\partial _{p}\right] _{p=0}^{n}\frac{Z\left[ 2p\right] }{Z\left[ p\right]
^{2}}+O\left( N^{-2}\right) .
\end{equation}

At this point we need to use a particular generating function $Z\left[
p\right] $. For a Gaussian variable with the generating function given by
Eq.~(\ref{gaussz}) the expectation value becomes
\begin{equation}
\label{chi1gauans}
\left\langle \hat{\chi }_{n}\right\rangle =\chi _{n}-\frac{N^{-1}}{2}\left[
i\partial _{p}\right] _{p=0}^{n}e^{-\sigma ^{2}p^{2}}+O\left( N^{-2}\right)
.
\end{equation}
This is the same as Eqs.~(\ref{chi1anse})-(\ref{chi1anso}).

Eq.~(\ref{chi1genans}) expresses the cumulant estimator bias for an
arbitrary distribution through its generating function $Z\left[ p\right] $.
Although it can be seen that the bias for a non-Gaussian distribution will
be different from that of Eq.~(\ref{chi1gauans}), we do not need its
general expression since we are only interested in testing the hypothesis
of an underlying Gaussian distribution.

The calculation of the covariances of cumulants (Eq.~(\ref{avchi2})) is a
little more involved since we need to expand both logarithmic terms prior
to taking the derivatives in $t$. We begin with one logarithmic term
{[}cf.~Eq.~(\ref{lnzoneterm}){]}, use the Gaussian generating function
$Z\left[ t\right] $ and evaluate the $n$-th derivative in $p$ while keeping
the $t$ variable intact,
\begin{eqnarray}
&  & \left[ i\partial _{p}\right] ^{n}\left( Z\left[ t\right] \ln Z\left[
p\right] +N^{-1/2}Z\left[ t\right] \exp \left( -\sigma ^{2}pt\right)
-\frac{N^{-1}}{2}Z\left[ t\right] \exp \left( -\sigma ^{2}p^{2}-2\sigma
^{2}pt\right) \right) \nonumber \\ & = & Z\left[ t\right] \chi
_{n}+N^{-1/2}Z\left[ t\right] \left( -i\right) ^{n}\sigma
^{2n}t^{n}-\frac{N^{-1}}{2}Z\left[ t\right] \left[ i\partial _{p}\right]
_{p=0}^{n}\exp \left( -\sigma ^{2}p^{2}-2\sigma ^{2}pt\right) .
\end{eqnarray}
(We shall not need the full expansion of the last cumbersome derivative.)
Then we act on this expression with another logarithmic term (from
Eq.~(\ref{logzte})), set $t=0$ and take the $m$-th derivative in $p$. After
some straightforward algebra we obtain Eq.~(\ref{chi2ans}):
\begin{equation}
\left\langle \hat{\chi }_{m}\hat{\chi }_{n}\right\rangle -\left\langle
\hat{\chi }_{m}\right\rangle \left\langle \hat{\chi }_{n}\right\rangle
=N^{-1}\left[ i\partial _{p}\right] _{p=0}^{m}\left( -i\right) ^{n}\sigma
^{2n}p^{n}=N^{-1}n!\sigma ^{2n}\delta _{mn}+O\left( N^{-2}\right) .
\end{equation}

The same method and Eq.~(\ref{chi1genans}) can be used to compute the
expectation values and covariances of cumulant estimators also for
non-Gaussian distributions as long as the generating function $Z\left[
p\right] $ is given. Expectation values of products of two or more
cumulants (e.g.~$\left\langle \hat{\chi }_{l}\hat{\chi }_{m}\hat{\chi
}_{n}\right\rangle $) can be found as well, although the calculations are
cumbersome. However, in the case of the underlying Gaussian distribution a
simple algebraic consideration shows that higher-order correlations between
the cumulant estimators $\hat{\chi }_{n}$ are of order $O\left(
N^{-2}\right) $ or smaller, and therefore $\hat{\chi }_{n}$ themselves can
be approximately treated as independent Gaussian variables. This is found
by noticing that the logarithmic derivative operator of Eq.~(\ref{logzte})
contains terms of order $O\left( N^{-1/2}\right) $ and $O\left(
N^{-1}\right) $ while we are only interested in the $O\left( N^{-1}\right)
$ terms in the final result. We could introduce a formal operator $L_{n}$
by
\begin{equation}
\label{eq:lndef}
L_{n}\equiv \left[ i\partial _{p}\right] ^{n}\ln \tilde{Z}\left[ p,t\right]
_{p=0}=\chi _{n}+N^{-1/2}A_{n}+N^{-1}B_{n}+O\left( N^{-3/2}\right) .
\end{equation}
The operator coefficients $A_{n}$ and $B_{n}$ will later act on $Z\left[
t\right] $ with their derivatives in $t$ evaluated only at the end; note
that $A_{n}Z\left[ t\right] _{t=0}=0.$ Then Eq.~(\ref{avchi2}) for the
correlation between the cumulant estimators is rewritten as
\begin{equation}
\left\langle \hat{\chi }_{m}\hat{\chi }_{n}\right\rangle =L_{m}L_{n}Z\left[
t\right] _{t=0}+O\left( N^{-3/2}\right) .
\end{equation}
Selecting the $O\left( N^{-1}\right) $ terms in the product gives
\begin{equation}
L_{m}L_{n}Z\left[ t\right] _{t=0}-\left( L_{m}Z\left[ t\right]
_{t=0}\right) \left( L_{n}Z\left[ t\right] _{t=0}\right)
=N^{-1}A_{m}A_{n}Z\left[ t\right] _{t=0}+...
\end{equation}
where all terms containing $B_{n}$ cancel. The only surviving $O\left(
N^{-1}\right) $ term came from the product of two $O\left( N^{-1/2}\right)
$ non-commuting operator terms, and all other terms contained a commuting
$\chi _{n}$ and canceled after subtracting the product of the expectation
values. Now we notice that the $O\left( N^{-1}\right) $ terms in a product
of more than two operators $L_{n}$ from Eq.~(\ref{eq:lndef}) must contain
at least one factor $\chi _{n}$. It follows that all ``connected
correlators'' of more than two cumulant estimators, such as
\begin{equation}
\left\langle \hat{\chi }_{l}\hat{\chi }_{m}\hat{\chi }_{n}\right\rangle
-\left\langle \hat{\chi }_{l}\right\rangle \left\langle \hat{\chi
}_{m}\hat{\chi }_{n}\right\rangle -\left\langle \hat{\chi
}_{m}\right\rangle \left\langle \hat{\chi }_{l}\hat{\chi }_{n}\right\rangle
-\left\langle \hat{\chi }_{n}\right\rangle \left\langle \hat{\chi
}_{l}\hat{\chi }_{m}\right\rangle +2\left\langle \hat{\chi
}_{l}\right\rangle \left\langle \hat{\chi }_{m}\right\rangle \left\langle
\hat{\chi }_{n}\right\rangle
\end{equation}
(for any $l$, $m$, $n$) contain no $O\left( N^{-1}\right) $
terms. Therefore the cumulant estimators $\hat{\chi }_{n}$ in the limit
of large number of samples $N$ are approximately Gaussian distributed
and independent variables (up to terms of order $O\left( N^{-2}\right) $).

\section{Properties of multivariate cumulants}

\label{app:mc}Here we examine the estimators $\hat{\chi }_{mn}$ of
cumulants of a distribution of two Gaussian variables $\left( x,y\right) $
which are independent and have variances $\sigma _{x}^{2}$ and $\sigma
_{y}^{2}$. We show that the cumulant estimators are approximately
independent in the limit of large sample size $N$, namely that their
covariances are
\begin{equation}
\left\langle \hat{\chi }_{kl}\hat{\chi }_{mn}\right\rangle -\left\langle
\hat{\chi }_{kl}\right\rangle \left\langle \hat{\chi }_{mn}\right\rangle
=N^{-1}m!n!\, \delta _{km}\delta _{ln}\sigma _{x}^{2m}\sigma
_{y}^{2n}+O\left( N^{-2}\right) ,
\end{equation}
while the means are biased by a quantity of order $N^{-1}$,
\begin{equation}
\left\langle \hat{\chi }_{kl}\right\rangle =O\left( N^{-1}\right) .
\end{equation}
Then we show how to generalize the results of the previous Appendix to
multivariate cumulants.

Cumulants of a distribution of two variables $\left( x,y\right) $ are
defined by analogy with Eq.~(\ref{cumu-gen}),
\begin{equation}
\label{eq:cumu-two-gen}
\chi _{mn}=\left[ i\partial _{p}\right] ^{m}\left[ i\partial _{q}\right]
^{n}\ln Z\left[ p,q\right] _{p=q=0},
\end{equation}
where $Z\left[ p,q\right] $ is the generating function of moments of the
distribution,
\begin{equation}
\label{eq:mom-two-gen}
\mu _{mn}\equiv \left\langle x^{m}y^{n}\right\rangle =\left[ i\partial
_{p}\right] ^{m}\left[ i\partial _{q}\right] ^{n}Z\left[ p,q\right]
_{p=q=0}.
\end{equation}
A Gaussian distribution is characterized by
\begin{equation}
Z\left[ p,q\right] =\exp \left[ -\frac{1}{2}\left( p,q\right) {\mathbf
C}\left( p,q\right) ^{T}\right]
\end{equation}
where we multiplied the row and column $2$-vectors $\left( p,q\right) $
by the appropriate $2\times 2$ correlation matrix ${\mathbf C}$. (This
matrix is diagonal if the variables $\left( x,y\right) $ are independent,
but we shall not need this for most of the derivation.)

We follow the same method of calculation as in Appendix A and introduce a
modified (operator-valued) generating function
\begin{equation}
\tilde{Z}\left[ p,q;t,u\right] \equiv Z\left[ p,q\right] +N^{-1/2}\exp
\left[ p\partial _{t}+q\partial _{u}\right]
\end{equation}
to be used instead of $\tilde{Z}\left[ p,t\right] $ in Eqs.~(\ref{avchi1}),
(\ref{avchi2}). We omit the calculations since they are very similar to
those in Appendix A and only cite the results. The bias in the cumulant
estimators $\hat{\chi }_{mn}$ is described by a formula analogous to
Eq.~(\ref{chi1gauans}),
\begin{equation}
\label{eq:chimnave}
\left\langle \hat{\chi }_{mn}\right\rangle =\chi
_{mn}-\frac{N^{-1}}{2}\left[ i\partial _{p}\right] ^{m}\left[ i\partial
_{q}\right] ^{n}\exp \left[ -\left( p,q\right) {\mathbf C}\left( p,q\right)
^{T}\right] _{p=q=0}.
\end{equation}

The covariance of two cumulant estimators $\hat{\chi }_{kl}$ and $\hat{\chi
}_{mn}$ vanishes unless the cumulants are of the same order, $k+l=m+n$, in
which case it is given by
\begin{equation}
\label{eq:chimn2ave}
\left\langle \hat{\chi }_{kl}\hat{\chi }_{mn}\right\rangle -\left\langle
\hat{\chi }_{kl}\right\rangle \left\langle \hat{\chi }_{mn}\right\rangle
=N^{-1}\left[ i\partial _{t}\right] ^{k}\left[ i\partial _{u}\right]
^{l}\left[ i\partial _{p}\right] ^{m}\left[ i\partial _{q}\right] ^{n}\exp
\left[ -\left( t,u\right) {\mathbf C}\left( p,q\right) ^{T}\right]
_{p=q=t=u=0}.
\end{equation}

In our case of interest, the covariance matrix ${\mathbf C}$ is diagonal,
\begin{equation}
C=\left( \begin{array}{cc}
\sigma _{x}^{2} & 0\\
0 & \sigma _{y}^{2}
\end{array}\right) ,
\end{equation}
and then Eqs.~(\ref{eq:chimnave})-(\ref{eq:chimn2ave}) simplify to
\begin{equation}
\label{eq:chimnpq-ave}
\left\langle \hat{\chi }_{mn}\right\rangle =\chi
_{mn}-\frac{N^{-1}}{2}\left[ i\partial _{p}\right] ^{m}\left[ i\partial
_{q}\right] ^{n}\exp \left[ -p^{2}\sigma _{x}^{2}-q^{2}\sigma
_{y}^{2}\right] _{p=q=0}+O\left( N^{-2}\right) ,
\end{equation}
\begin{equation}
\label{eq:chimnqp-cov}
\left\langle \hat{\chi }_{kl}\hat{\chi }_{mn}\right\rangle -\left\langle
\hat{\chi }_{kl}\right\rangle \left\langle \hat{\chi }_{mn}\right\rangle
=N^{-1}m!n!\delta _{km}\delta _{ln}\sigma _{x}^{2m}\sigma _{y}^{2n}+O\left(
N^{-2}\right) .
\end{equation}

The expressions of Eqs.~(\ref{eq:chimnpq-ave})-(\ref{eq:chimnqp-cov}) are
straightforwardly generalized for distributions of three and more
variables, for example
\begin{equation}
\textrm{var}\left[ \hat{\chi }_{klm}\right] =N^{-1}k!l!m!\sigma
_{x}^{2k}\sigma _{y}^{2l}\sigma _{z}^{2m}+O\left( N^{-2}\right) .
\end{equation}

For the case of cumulants of Fourier components of a Gaussian random field,
the individual Fourier modes are independently distributed with variances
equal to the power spectrum $P\left( k\right) $. Therefore the expressions
we derived are applicable directly with the substitution $\sigma
_{x}^{2}=P\left( k_{1}\right) $, $\sigma _{y}^{2}=P\left( k_{2}\right)
$,... for the appropriate modes. A general Fourier cumulant
$\tilde{C}^{\left( n\right) }\left( {\mathbf k}_{1},...,{\mathbf
k}_{n}\right) $ of Eq.~(\ref{eq:foucumz}) with all $n$ vectors ${\mathbf
k}_{i}$ distinct will be estimated from $N$ samples obtained by rotating
the set of vectors $\left\{ {\mathbf k}_{i}\right\} $ and is in our
notation a cumulant $\chi _{11...1}$ of the $n$-variable distribution of
the $n$ Fourier modes $a_{{\mathbf k}_{i}}$; its sample variance is
therefore
\begin{equation}
\label{eq:varcgen}
\textrm{var}\left[ \tilde{C}^{\left( n\right) }\left( {\mathbf
k}_{1},...,{\mathbf k}_{n}\right) \right] =N^{-1}P\left( k_{1}\right)
...P\left( k_{n}\right) +O\left( N^{-2}\right) .
\end{equation}

\section{Generating functionals for random superpositions of shapes}

\label{app:gfs}Here we derive the generating functional for the random
field resulting from a superposition of fixed profiles (shapes), centered
at points (``seeds'') drawn from some known distribution and scaled and
rotated randomly. In particular, we show how to use the generating
functional to obtain the cumulants of Fourier components for such a random
field. We first consider the simpler case of the Poisson distribution of
seeds and then briefly show how to generalize the same formalism for
non-Poisson distributions of seed centers.

We work in flat space for simplicity; although we only need the result in
two dimensions, our derivation applies to any $d$-dimensional space. We
consider a finite region of space with unit area, so that $\int dx=1$. If
we start with a shape of a given profile $s\left( x\right) $ and translate
it to a set of locations $x_{1}$, ..., $x_{n}$, the result is
\begin{equation}
\label{f-def}
f\left( x;x_{1},...,x_{n}\right) =\sum _{k}s\left( x-x_{k}\right) .
\end{equation}
To compute the generating functional for the distribution $f\left(
x;x_{1},...,x_{n}\right) $ we need to average the following over all
numbers $n$ of centers and center positions $x_{k}$:
\begin{equation}
\label{zj-simple}
Z\left[ J\left( x\right) \right] =\left\langle \exp \left( -i\int f\left(
x;x_{1},...,x_{n}\right) J\left( x\right) dx\right) \right\rangle .
\end{equation}
The centers $x_{i}$ are Poisson distributed with a fixed mean number of
centers $n_{c}$ in the whole region, and the probability of having $n$
centers is $\left( n_{c}\right) ^{n}\exp \left( -n_{c}\right) /n!$. The
averaging in Eq.~(\ref{zj-simple}) with $f\left( x\right) $ defined by
Eq.~(\ref{f-def}) then gives
\begin{eqnarray}
Z\left[ J\left( x\right) \right]  & = & e^{-n_{c}}\sum ^{\infty
}_{n=0}\frac{n_{c}^{n}}{n!}\int dx_{1}...dx_{n}\exp \left( -i\int f\left(
x;x_{1},...,x_{n}\right) J\left( x\right) dx\right) \nonumber \\ & = & \exp
\left( -n_{c}+n_{c}\int dx_{0}\exp \left( -i\int s\left( x-x_{0}\right)
J\left( x\right) dx\right) \right) \label{eq:zspoi-simple}
\end{eqnarray}
As usual, the logarithm $\ln Z\left[ J\left( x\right) \right] $ of the
generating functional generates the cumulants $C^{\left( n\right) }\left(
x_{1},...,x_{n}\right) $ of the distribution,
\begin{equation}
\ln Z\left[ J\left( x\right) \right] =\int C^{\left( 1\right) }\left(
x_{1}\right) \frac{J\left( x_{1}\right) }{i}dx_{1}+\int C^{\left( 2\right)
}\left( x_{1},x_{2}\right) \frac{J\left( x_{1}\right) J\left( x_{2}\right)
}{i^{2}2!}dx_{1}dx_{2}+...
\end{equation}
(this can also be taken as a definition of cumulants) and therefore is of
most interest for us.

We now add more variety to the random field $f\left(
x;x_{1},...,x_{n}\right) $ by allowing the shapes $s\left( x\right) $ to be
rotated, scaled and attenuated. The rotation is effected by introducing an
orthogonal matrix $R$ uniformly distributed in the orthogonal group of
$d$-dimensional rotations ${\mathbf O}\left( d\right) $; the scaling by a
factor $\lambda $ distributed according to some probability density
$p_{\lambda }\left( \lambda \right) $; and the attenuation by multiplying
the profile by an overall factor $\nu $ with probability density $p_{\nu
}\left( \nu \right) $. We give each shape an individual randomly selected
rotation angle, scale and attenuation. The transformed profile is $\nu
s\left( \lambda ^{-1}R\left( x-x_{0}\right) \right) $ and the generating
functional is similar to the one above:
\begin{eqnarray}
\ln Z\left[ J\left( x\right) \right] =-n_{c}+n_{c}\int dx_{0}dR\,
p_{\lambda }\left( \lambda \right) d\lambda \, p_{\nu }\left( \nu \right)
d\nu  &  & \nonumber \\ \times \exp \left( -i\nu \int s\left( \lambda
^{-1}R\left( x-x_{0}\right) \right) J\left( x\right) dx\right) . &  &
\label{lnz-gen-r}
\end{eqnarray}
Here the integration measure $dR$ for rotations is assumed to be normalized
to unity, as well as with all other integrations. We shall below drop the
irrelevant additive constant $n_{c}$ in Eq.~(\ref{lnz-gen-r}).

The non-Gaussian components of the distribution are now read from
Eq.~(\ref{lnz-gen-r}). After the exponential is expanded in powers of
$J\left( x\right) $ under the integral, the general cumulant of order $n$
is formed from the terms of order $J\left( x\right) ^{n}$ and can be
readily found for any given profile $s\left( x\right) $ and any assumed
distributions of scaling and attenuation. The resulting non-Gaussian
distribution is homogeneous and isotropic due to averaging over spatial
position $x_{0}$ and the rotations $R$.

To illustrate the method, we extract from Eq.~(\ref{lnz-gen-r}) the
characteristic function $\tilde{f}\left( j\right) $ of the one-point
distribution of $f\left( x;x_{1},...,x_{n}\right) $ at a fixed reference
point $x=x_{*}$. This is done by substituting $J\left( x\right) =j\delta
\left( x-x_{*}\right) $ into Eq.~(\ref{lnz-gen-r}); the result is the
generating function of cumulants for the one-point distribution,

\begin{equation}
\ln \tilde{f}\left( j\right) \equiv \sum _{n}\frac{j^{n}}{i^{n}n!}\chi
_{n}=n_{c}\left\langle \lambda \right\rangle \int dx\, p_{\nu }\left( \nu
\right) d\nu \, \exp \left( -i\nu s\left( x\right) j\right)
\end{equation}
and the $n$-th cumulant $\chi  _{n}$ of that distribution is
\begin{equation}
\chi _{n}=n_{c}\left\langle \lambda \right\rangle \left\langle \nu
^{n}\right\rangle \left\langle s^{n}\left( x\right) \right\rangle _{x}.
\end{equation}

In the same manner, one can obtain the generating functional for the
distribution of the Fourier components of $f\left( x\right) $. By Fourier
transforming Eq.~(\ref{lnz-gen-r}) we obtain
\begin{equation}
\label{eq:zjk}
\ln Z\left[ j\left( k\right) \right] =n_{c}\int dx\, dR\, p_{\lambda
}\left( \lambda \right) d\lambda \, p_{\nu }\left( \nu \right) d\nu \, \exp
\left( -i\nu \int \tilde{s}\left( -\lambda Rk\right) j\left( k\right)
e^{-ikx}\frac{dk}{\left( 2\pi \right) ^{d}}\right) .
\end{equation}
The $n$-th Fourier space cumulant $\tilde{C}^{\left( n\right) }\left(
k_{1},...,k_{n}\right) $ is then given by
\begin{equation}
\label{c-f-gen}
\tilde{C}^{\left( n\right) }\left( k_{1},...,k_{n}\right)
=n_{c}\left\langle \nu ^{n}\right\rangle \int dR\, p_{\lambda }\left(
\lambda \right) d\lambda \, \delta \left( k_{1}+...+k_{n}\right) \prod
^{n}_{i=1}\tilde{s}\left( -\lambda Rk_{i}.\right)
\end{equation}
Note that since $\tilde{s}\left( k\right) $ is a Fourier transform of a
real profile function $s\left( x\right) $ and the rotations $R$ include
mirror symmetry $k\rightarrow -k$, the cumulant of Eq.~(\ref{c-f-gen})
is always real-valued.

Although Eq.~(\ref{c-f-gen}) shows that cumulants of all orders are
generally expected to be nonzero, we need to estimate the statistical
significance of their deviation from zero compared to the sample variance.
The variance of a cumulant estimator, as derived above, is proportional to
the appropriate powers of the power spectrum. The power spectrum $P\left(
k\right) $ of the seed-induced distribution can be found from
Eq.~(\ref{c-f-gen}) as the second-order cumulant $\tilde{C}^{\left(
2\right) }\left( k,-k\right) $,
\begin{equation}
\label{eq:pksgen}
P\left( k\right) =n_{c}\left\langle \nu ^{2}\right\rangle \int dR\,
p_{\lambda }\left( \lambda \right) d\lambda \, \left| \tilde{s}\left(
-\lambda Rk\right) \right| ^{2}.
\end{equation}
To understand the qualitative behavior of the cumulant estimators, we shall
simplify the case by assuming that the distribution of scales $p_{\lambda
}\left( \lambda \right) $ is trivial and that the profile $s\left( x\right)
$ is spherically symmetric. In that case, the integrals in
Eqs.~(\ref{c-f-gen}), (\ref{eq:pksgen}) simplify
\begin{equation}
\tilde{C}^{\left( n\right) }\left( k_{1},...,k_{n}\right)
=n_{c}\left\langle \nu ^{n}\right\rangle \tilde{s}\left( k_{1}\right)
...\tilde{s}\left( k_{n}\right) ,
\end{equation}
\begin{equation}
P\left( k\right) =n_{c}\left\langle \nu ^{2}\right\rangle \left|
\tilde{s}\left( k\right) \right| ^{2}.
\end{equation}
If the cumulant $\tilde{C}^{\left( n\right) }\left( k_{1},...,k_{n}\right)
$ is estimated using a sample of $N$ values, the ratio of the expected
signal to the standard deviation of the estimator for a Gaussian sample
(assuming all $k_{i}$ are different) is
\begin{equation}
\label{eq:cn-sn}
\frac{\tilde{C}^{\left( n\right) }\left( k_{1},...,k_{n}\right)
}{\sqrt{\textrm{var}\left[ \tilde{C}^{\left( n\right) }\left(
k_{1},...,k_{n}\right) \right]
}}=\frac{\sqrt{N}}{n_{c}^{n/2-1}}\frac{\left\langle \nu ^{n}\right\rangle
}{\left\langle \nu ^{2}\right\rangle ^{n/2}}\frac{\tilde{s}\left(
k_{1}\right) ...\tilde{s}\left( k_{n}\right) }{\left| \tilde{s}\left(
k_{1}\right) \right| ...\left| \tilde{s}\left( k_{n}\right) \right| }.
\end{equation}
The last ratio of various $\tilde{s}\left( k\right) $ in
Eq.~(\ref{eq:cn-sn}) is equal to $1$ if the homogeneity constraint
$k_{1}+...+k_{n}=0$ is satisfied; also the ratio containing $\nu $ should
be of order $1$ except for specially engineered attenuation distributions.
We obtain therefore the general result that the sensitivity of an
individual cumulant estimator is decreased with the cumulant order $n$ and
with the expected number $n_{c}$ of seeds in the observation region (the
latter is a manifestation of the central limit theorem), and grows with the
number $N$ of sample points which is determined by the number of modes
$a_{{\mathbf k}}$ at the chosen scales as $\sqrt{N}$. We stress that
Eq.~(\ref{eq:cn-sn}) provides only an estimate of the sensitivity under
simplifying assumptions of spherical symmetry and fixed scale of seed
profiles.

Finally, we generalize our formalism to include arbitrary (non-Poisson)
distributions of seed centers. Similarly to a continuous random field, a
distribution of seed centers can be fully specified by a suitable
generating functional. Constructing it is equivalent to specifying all the
connected correlation functions $\xi \left( x_{1},...,x_{n}\right) $ of
seed positions. Assume that we are given all joint probability densities
$Prob\left( x_{1},...,x_{n}\right) \prod _{k}dx_{k}$ for having a seed at
each of the positions $x_{1}$,...,$x_{n}$. If we denote the one-point seed
density $Prob(x_{1})\equiv n_{s}\left( x_{1}\right) $, then the two-point
seed correlation function $\xi \left( x_{1},x_{2}\right) $ is usually
defined by
\begin{equation}
\label{eq:prob12xi}
Prob(x_{1},x_{2})\equiv n_{s}\left( x_{1}\right) n_{s}\left( x_{2}\right)
\left[ 1+\xi \left( x_{1},x_{2}\right) \right] .
\end{equation}
The three-point function $\xi \left( x_{1},x_{2},x_{3}\right) $ is then
defined from the relation

\begin{eqnarray}
Prob\left( x_{1},x_{2},x_{3}\right) \equiv n_{s}\left( x_{1}\right)
n_{s}\left( x_{2}\right) n_{s}\left( x_{3}\right) \left[ 1+\xi \left(
x_{1},x_{2}\right) \right.  &  & \nonumber \\ \left. +\xi \left(
x_{1},x_{3}\right) +\xi \left( x_{2},x_{3}\right) +\xi \left(
x_{1},x_{2},x_{3}\right) \right]  &  & \label{eq:prob123xi}
\end{eqnarray}
and similarly for the higher-order correlations. The relation between the
joint probability densities $Prob\left( x_{1},...,x_{n}\right) $ and the
connected correlation functions $\xi \left( x_{1},...,x_{n}\right) $ is
similar to that of moments and cumulants of a continuous random field
(except that $\xi $'s in our case are multiplied by several factors of
$n_{s}$). We can define the generating functional for the seed distribution
$Z_{s}\left[ J\left( x\right) \right] $ as the functional that generates
the joint probabilities,
\begin{equation}
\label{eq:genseed}
Z_{s}\left[ J\left( x\right) \right] \equiv 1+\sum _{n=1}^{\infty }\int
dx_{1}...dx_{n}\frac{J\left( x_{1}\right) ...J\left( x_{n}\right)
}{i^{n}n!}Prob\left( x_{1},...,x_{n}\right) .
\end{equation}

One can see from Eqs.~(\ref{eq:prob12xi}), (\ref{eq:prob123xi}) that the
connected correlation functions $\xi \left( x_{1},x_{2},...\right) $ are
generated by the logarithm of $Z_{s}$, similarly to the usual cumulants
except for the normalization to the one-point density $n_{s}$:
\begin{equation}
\label{eq:genseedc}
\xi \left( x_{1},...,x_{n}\right) =i^{n}\frac{\delta ^{n}}{\delta J\left(
x_{1}\right) ...\delta J\left( x_{n}\right) }\ln Z_{s}\left[
\frac{1}{n_{s}\left( x\right) }J\left( x\right) \right] _{J=0}.
\end{equation}
Conversely, if we are given the full set of connected correlators $\xi
\left( x_{1},...,x_{n}\right) ,\, n\geq 2$, we use Eq.~(\ref{eq:genseedc})
to construct the generating functional for seeds $Z_{s}\left[ J\right] $
(we would need to formally introduce the one-point correlators as $\xi
\left( x_{1}\right) \equiv 1$ to use Eq.~(\ref{eq:genseedc}) with $n=1$ and
also fix $Z_{s}\left[ J=0\right] =1$).

A simple example is a Poisson distribution where the seeds are completely
uncorrelated, $\xi \left( x_{1},...,x_{n}\right) =0$ for $n\geq 2$ and the
distribution is completely specified by the seed density $n_{s}\left(
x\right) $. We obtain a generating functional
\begin{equation}
\label{eq:zspoi}
Z_{s}^{(\textrm{P})}\left[ J\left( x\right) \right] =\exp \left[ -i\int
n_{s}\left( x\right) J\left( x\right) dx\right] .
\end{equation}

It turns out that given any generating functional of seeds $Z_{s}\left[
J\right] $, one can directly obtain the generating functional $Z\left[
J\right] $ for the ``seeded'' random field of Eq.~(\ref{f-def}). To show
this, we compare the definitions of both functionals. The definition of
$Z\left[ J\right] $ in Eq.~(\ref{zj-simple}) is the average of the quantity
\begin{equation}
\exp \left( -i\int f\left( x;x_{1},...,x_{n}\right) J\left( x\right)
dx\right) \equiv \prod _{k=1}^{n}E\left( x_{k};J\right) ,
\end{equation}

\begin{equation}
\label{eq:zfzaux}
E\left( x;J\right) \equiv \exp \left[ -i\int J\left( x^{\prime }\right)
s\left( x^{\prime }-x\right) dx^{\prime }\right] ,
\end{equation}
over all seed numbers $n$ and seed positions $x_{k}$. The average
is weighed by the probabilities $P\left( x_{1},...,x_{n}\right) \prod
_{k}dx_{k}$ of having seeds at points $x_{1}$,...,$x_{n}$ and nowhere else.
These probabilities are related to the probability densities $Prob\left(
x_{1},...,x_{n}\right) $ by
\begin{equation}
P\left( x_{1},...,x_{n}\right) =P\left( \emptyset \right) Prob\left(
x_{1},...,x_{n}\right)
\end{equation}
where $P\left( \emptyset \right) =\exp \left[ -\int n_{s}\left( x\right)
dx\right] $ is the (finite) probability of having no seeds anywhere. Then
Eq.~(\ref{zj-simple}) can be rewritten as
\begin{eqnarray}
Z\left[ J\left( x\right) \right] =P\left( \emptyset \right) \left[ 1+\int
dx_{1}Prob\left( x_{1}\right) E\left( x_{1};J\right) \right.  &  & \\
\left. +\int \frac{dx_{1}dx_{2}}{2!}Prob\left( x_{1},x_{2}\right) E\left(
x_{1};J\right) E\left( x_{2};J\right) +...\right]  &  & \label{eq:zjrewr}
\end{eqnarray}
We notice that Eqs.~(\ref{eq:genseed}) and (\ref{eq:zjrewr}) are
essentially the same, and therefore
\begin{equation}
\label{eq:zfzsans}
Z\left[ J\left( x\right) \right] =P\left( \emptyset \right) Z_{s}\left[
iE\left( x;J\right) \right] .
\end{equation}

Eq.~(\ref{eq:zfzsans}) is the main relation which allowed us above to
obtain the analytic generating functional for the Poisson distribution of
seeds, cf. Eqs.~(\ref{eq:zspoi-simple}), (\ref{eq:zspoi}) with $n_{s}\left(
x\right) =const$. The generating functional $Z\left[ j\left( k\right)
\right] $ for the Fourier modes of the random field is obtained in a
similar way,
\begin{equation}
Z\left[ j\left( k\right) \right] =P\left( \emptyset \right) Z_{s}\left[
i\tilde{E}\left( x;j\right) \right] ,\quad \tilde{E}\left( {\mathbf
x};j\left( {\mathbf k}\right) \right) \equiv \exp \left[ -i\int j\left(
{\mathbf k}\right) \tilde{s}\left( {\mathbf k}\right) e^{-i{\mathbf
kx}}d{\mathbf k}\right] .
\end{equation}
These simple relations are quite general and allow to compute the
generating functional $Z\left[ j\left( k\right) \right] $ for arbitrary
distribution of seeds given by a generating functional $Z_{s}\left[ J\left(
x\right) \right] $.

\end{document}